\documentclass[11pt]{article}

\usepackage{amsmath,amssymb}
\usepackage{slashed}
\usepackage{xcolor} % Define colors
\usepackage{graphicx}
\usepackage{dcolumn} % Align table columns on decimal point
\usepackage{bm} % Bold math
\usepackage{epsfig}
\usepackage{epstopdf}
\usepackage{grffile}
\usepackage{color}
\usepackage{colordvi}
\usepackage{amsmath,amssymb}
\usepackage{rotating}
\usepackage{lscape}
\usepackage{cite}
\usepackage{float}
\usepackage{hyperref}

\usepackage [latin1]{inputenc}

\def\Re{{\cal R \mskip-4mu \lower.1ex \hbox{\it e}\,}}
\def\Im{{\cal I \mskip-5mu \lower.1ex \hbox{\it m}\,}}

\def\tev{\,{\ifmmode\mathrm {TeV}\else TeV\fi}}
\def\gev{\,{\ifmmode\mathrm {GeV}\else GeV\fi}}
\def\mev{\,{\ifmmode\mathrm {MeV}\else MeV\fi}}

\begin{document}

%%%%%%%%%%%%%%%%%%%%%%%%%%%%%%

\begin{center}

\vspace*{15mm}
\vspace{1cm}
{\Large \bf  Constraining the top quark effective field theory using the top quark pair production in association with a jet at future lepton colliders}

\vspace{1cm}

{\small \bf  Reza Jafari$^{1}$, Parvin Eslami$^{1}$,  Mojtaba Mohammadi Najafabadi$^{2}$, Hamzeh Khanpour$^{3,2}$ }

 \vspace*{0.5cm}

{\small\sl$^{1}$Department of Physics, Faculty of Science, Ferdowsi University of Mashhad, Mashhad, Iran  } \\
{\small\sl$^{2}$School of Particles and Accelerators, Institute for Research in Fundamental Sciences (IPM) P.O. Box 19395-5531, Tehran, Iran } \\
{\small\sl$^{3}$Department of Physics, University of Science and Technology of
Mazandaran, P.~O.~Box 48518-78195, Behshahr, Iran }

\vspace*{.2cm}
\end{center}

\vspace*{10mm}

%
%%%%%%%%%%%%%%%%%%%%%%%%%%%%%%%%%%%%%%%%%%    abstract    %%%%%%%%%%%%%%%%%%%%%%%%%%%%%%%%%%%%%%%%%%%%%%%%%
%
\begin{abstract}\label{abstract}
{
Our main aim in this paper is to constrain the effective field theory describing the top quark 
couplings through the $e^{-} e^{+} \rightarrow t \bar{t}+$jet process. 
The analysis is carried out considering two different center-of-mass energies of 
500 and 3000 GeV including a realistic simulation of the detector response and the main sources of background processes. 
The expected limits at 95\% CL are derived on the new physics couplings 
such as $t \bar t \gamma$, $t \bar t Z$, and $t \bar t g$  for each benchmark scenario using the dileptonic $t \bar{t}$ final state.
We show that the 95\% CL limits on dimensionless Wilson coefficients $\bar{c}_i$ considered in this analysis could be probed
down to $10^{-4}$. 
%Our findings indicate that  including the top quark pair production in association with a jet to the $t\bar{t}$ process at future lepton colliders
%improves the constraints on the Wilson coefficients remarkably. 
}

\end{abstract}

\vspace*{3mm}

%{\bf Keywords}: Top quark, Beyond the Standard Model, Lepton Collider.

\newpage

%
%%%%%%%%%%%%%%%%%%%%%%%%%%%%%%%%%%%%%%%%%%    Introduction    %%%%%%%%%%%%%%%%%%%%%%%%%%%%%%%%%%%%%%%%%%%%%%%%%
\section{Introduction}\label{sec:intro}
{
Since the Higgs boson observation~\cite{Aad:2012tfa,Chatrchyan:2012xdj} at the Large Hadron Collider (LHC) 
by the ATLAS and CMS Collaborations, the primary focus of high energy particle physics is to probe 
its properties in details~\cite{Dawson:2013bba,deFlorian:2016spz,Heinemeyer:2013tqa}. 
In addition to the Higgs boson, the heaviest discovered particle to date, {\it i.e.} the top quark 
which was discovered by D0 and CDF Collaborations at Fermilab~\cite{Abe:1995hr,Abachi:1994td}, 
is expected to play an important role in the electroweak symmetry breaking (EWSB) mechanism due to its large mass.

Looking further into the future, in addition to the LHC, precision measurements of the top quark and Higgs boson properties  
motivate the construction of future lepton colliders which provide cleaner
environments due to the absence of hadronic initial state and their relatively smaller experimental uncertainties with respect to the hadron colliders.
Hence, there is currently a growing  interest in studying physics accessible by possible future high-energy and high-luminosity 
electron-positron colliders that would continue the investigations made with the large electron-positron (LEP)
collider to much higher energy and luminosity~\cite{Fan:2014vta,Moortgat-Picka:2015yla,Fujii:2015jha}. 
So far, there have been several proposals over the past  years for the future electron-positron colliders, 
such as the Compact Linear Collider (CLIC)~\cite{Charles:2018vfv,Roloff:2018dqu,Aicheler:2019dhf}, 
the International Linear Collider (ILC)~\cite{Aihara:2019gcq,Baer:2013cma,Behnke:2013xla}, 
Circular Electron Positron Collider (CPEC)~\cite{CEPC-SPPCStudyGroup:2015csa,CEPC-SPPCStudyGroup:2015esa}, 
and  the highest-luminosity energy frontier Future Circular Collider with electron-positron collisions
(FCC-ee) at CERN~\cite{Benedikt:2018qee}, previously known as TLEP~\cite{Gomez-Ceballos:2013zzn} 
(see, for example, the most recent Conceptual Design Report by FCC Collaboration~\cite{Abada:2019lih,Abada:2019zxq} for recent review.).

Phenomenological and experimental studies over the past decades have provided important information 
on the validity of the Standard Model (SM) as well as the physics beyond the 
SM (BSM)~\cite{Azzi:2019yne, Beacham:2019nyx,CidVidal:2018eel}. 
The focus of many studies has been on the top quark and Higgs boson phenomenology and search for the new physics  through them. 
These include, for example, the precise measurements of the top quark and Higgs boson masses, their couplings to the
other fundamental particles in the framework of the SM and BSM, and searches for new physics effects beyond the 
SM in both model dependent and independent ways. 
In the case that the possible new degrees of freedom are not light enough to be directly produced at a collider,
they could affect the SM observables indirectly through virtual effects.
In such conditions, a powerful tool to parametrise any potential deviations from the SM predictions
in a model-independent way is the standard model effective field theory (SMEFT).  
SMEFT  provides a general framework where non-redundant bases of independent operators can be built and one would be able
to  match them to explicit ultraviolet complete (UV-complete) models in a systematic way. 
From the phenomenological point of view, there is a large volume of published works to study  
the SMEFT in particular in the top quark and Higgs boson sectors from the LHC,  from electron-positron colliders, 
and from future proposed high-energy lepton-hadron and hadron-hadron colliders~\cite{e2,e3,e4,e5,e6,e7,e8,e9,e10,e11,Khanpour:2017cfq,	Hesari:2018ssq,Khanpour:2017inb,Ellis:2015sca,Chiu:2017yrx,Ellis:2017kfi,	Brooijmans:2016vro,Rontsch:2015una,Rontsch:2014cca,r1,r2,r3,r4,r5,r6,r7,r8,r9,r10,r11,Hartmann:2016pil,Fichet:2016iuo,Ellis:2014jta,Berthier:2015gja,Englert:2015hrx,Ellis:2014dva,
Denizli:2017pyu,Denizli:2019ijf,Hesari:2018lzx,ddd , Durieux:2018tev}.

The aim of the present study is to examine the sensitivity of the top quark pair production in association with a
jet at future electron-positron colliders to the SMEFT.  All dimension-six operators in the SILH basis
which involve top quark and/or Higgs and gauge bosons assuming CP-conservation are included~\cite{Artoisenet:2013puc,Alloul:2013naa}.
It is notable that flavour universality is assumed in the SILH basis.

We perform detailed sensitivity studies and present the expected 95\% CL limits on the operator coefficients for the center-of-mass 
energies of 500 and 3000 GeV with integrated luminosities $\cal L$ related to the proposed electron-positron colliders.
It is shown that including the $e^{-} e^{+} \rightarrow t \bar{t}+$jet process to $e^{-} e^{+} \rightarrow t \bar{t}$,  improves the 
sensitivity to the effective couplings of top quark with the electroweak gauge bosons.

This paper is organised as follows:
In section~\ref{sec:Theoretical}, the SMEFT framework is briefly introduced.
In section~\ref{sec:Simulation}, the details of the simulation for probing SMEFT operators through the production processes
of $t \bar{t}$ in association with a jet at the electron-positron collision are described.
In section~\ref{sec:Results}, the methodology applied in this analysis to constrain the Wilson coefficients, as well as the results, are presented. 
Finally, section~\ref{sec:Discussion} concludes the paper.   
}

%
%%%%%%%%%%%%%%%%%%%%%%%%%%%%%%%%%%%%%%%%%%        %%%%%%%%%%%%%%%%%%%%%%%%%%%%%%%%%%%%%%%%%%%%%%%%%
\section{ Theoretical framework}\label{sec:Theoretical}
{

As no clear evidence of new physics beyond the standard model has been observed, 
an efficient approach for examining the SM and possible deviations from SM could be provided by the SMEFT. 
In this approach, beyond the SM
effects are probed via a series of higher dimensional SM operators. 
The coefficients of the operators, so-called Wilson coefficients,  
can be connected to the parameters of explicit models.
The effective Lagrangian is provided considering  the operators which are invariant under the 
${\rm SU}(3) \times {\rm SU}(2) \times {\rm U}(1)$ gauge symmetries and Lorentz transformations.
We restrict ourselves to the operators with a lepton and baryon number conservation. In such a case, the leading contributions come
from dimension-six operators.
The general Lagrangian of the SM effective theory with dimension-six operators
is given by~\cite{Buchmuller:1985jz,Grzadkowski:2010es,Hagiwara:1993ck}, 

%-------------------------------------------
\begin{eqnarray}\label{eq:SMEFT}
{\cal L}_{\rm SMEFT} = {\cal L}_{\rm SM} + \sum_{i} \frac{c_{i} {\cal O}_{i}}{\Lambda^{2}} ,
\end{eqnarray}
%-------------------------------------------

In the above relation, $\Lambda$ is the energy scale of  new physics, $c_{i}$'s are  
dimensionless Wilson coefficients and the gauge-invariant dimension-six operators denoted by  ${\cal O}_{i}$ are constructed out of the SM fields.
There are various bases where the operators ${\cal O}_{i}$ are classified in an independent way.
In this work, the dimension-six operators sensitive to the $e^{-} e^{+} \rightarrow  t \bar{t}+$jet process are discussed
in the SILH basis \cite{Buchalla:2014eca, Buchalla:2015wfa, Contino:2013kra,  Artoisenet:2013puc}. 
This basis is not a unique basis and can be connected to the other bases.
The SMEFT Lagrangian $\cal L_{\rm SMEFT}$ in the SILH basis can be expressed as follows:

%-------------------------------------------
\begin{eqnarray}\label{eq:SMEFT-2}
&& {\cal L}_{\rm SMEFT} =  {\cal L}_{\rm SM} + {\cal L}_{\rm SILH} + {\cal L}_{F_1} + {\cal L}_{F_2} + {\cal L}_{\rm G}  + {\cal L}_{\rm CP}  \,.
\end{eqnarray}
%-------------------------------------------

The first term in the above effective Lagrangian, $ {\cal L}_{\rm SM}$,  is the well-known SM Lagrangian. 
The second term, ${\cal L}_{\rm SILH}$,  consists of a set of operators which involve the Higgs doublet $\Phi$
and could arise from UV-models where Higgs boson contributes to the strongly interacting sector. 
The interactions between two Higgs boson fields and a pair of quarks or a pair of leptons are described by ${\cal L}_{F_1}$
while  the interactions of a quark pair or a lepton
pair with one single Higgs field and a gauge boson are addressed by ${\cal L}_{F_2}$.
All the modifications related to the gauge sector, from  the gauge
bosons self energies to the gauge bosons self-interactions are parameterized in ${\cal L}_{\rm G}$.
The CP-violating interactions are described by $ {\cal L}_{\rm CP}$. In this work, the concentration is
on the CP-conserving operators.

Within the SMEFT framework, in addition to the new Feynman diagrams contributing to the $e^{-} e^{+} \rightarrow t \bar{t}+$jet,
the SM Feynman diagrams are modified. The representative
Feynman diagrams for  the top quark pair and $t\bar{t}(+$jet) production at electron-positron colliders are depicted in Fig. \ref{fig:feynman_diagrams_1}.
The filled circles are the vertices that receive modification from the SM effective field
theory. It is notable that in addition
to the diagrams for the $t\bar{t}g$ production where the SM couplings are modified, a new diagram
arising from $hgt\bar{t}$ contribute to the $e^{-} e^{+} \rightarrow t \bar{t} +$jet process. The contribution
of this diagram is small due to the presence of a Higgs boson Yukawa coupling with electron.

%------------------------------------------------
\begin{figure*}[htb]
	\vspace{0.50cm}	
		\begin{center}
	\resizebox{0.60\textwidth}{!}{\includegraphics{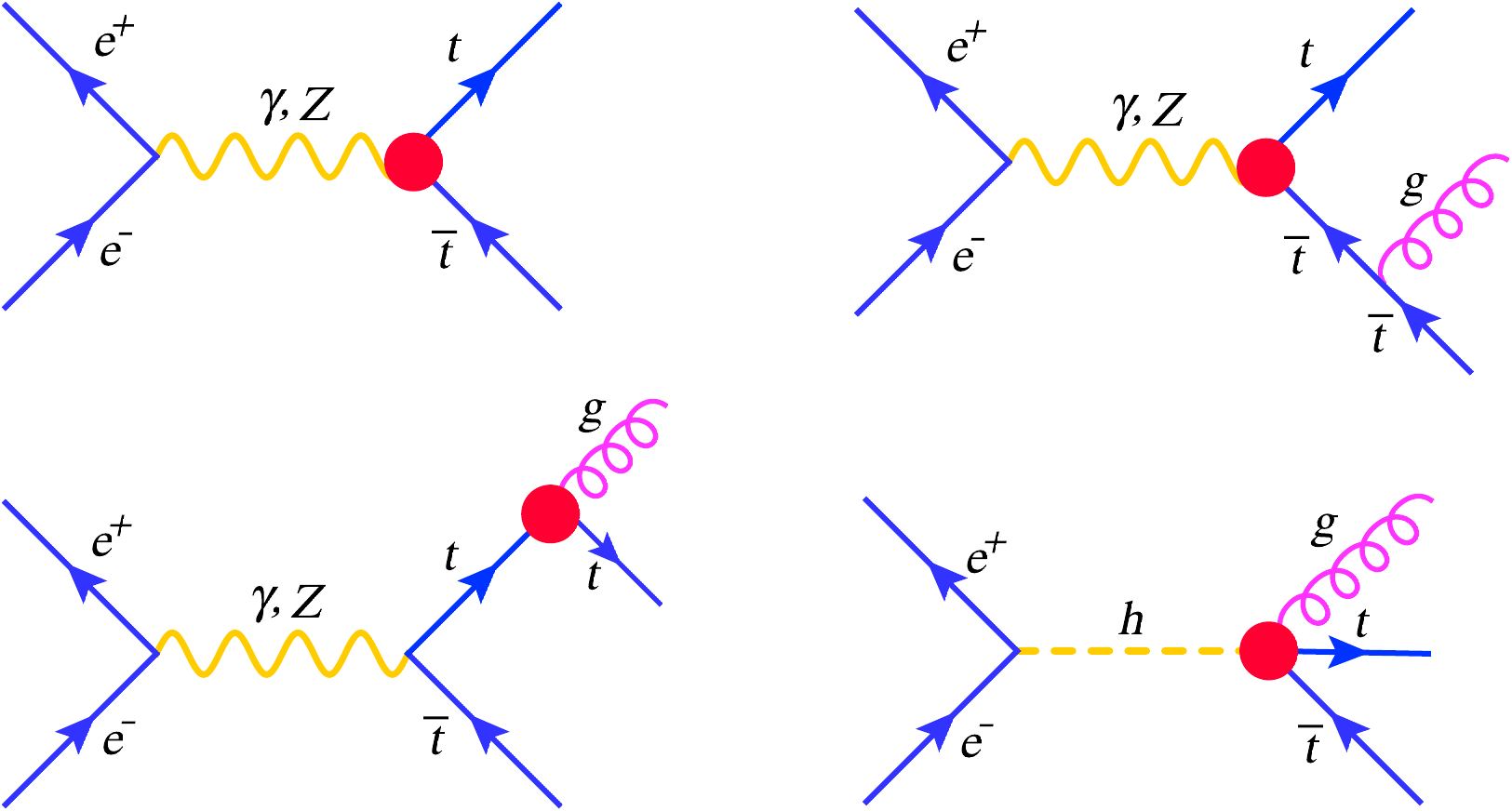}}
		\caption{{\small  Representative Feynman diagrams for the $t\bar{t}$ and $t\bar{t}$ production 
		in association with a jet in electron-positron collisions in the SMEFT. 
		 } \label{fig:feynman_diagrams_1}}
	\end{center}
\end{figure*}
%------------------------------------------------

In the current study, we restrict ourselves to effective operators contributing to $e^{-} e^{+} \rightarrow t \bar{t} +\text{jet}$ process
involving at least one top quark. Although other effective operators can affect 
the $t \bar{t} + \text{jet}$ process via for example  $Zee$ or $\gamma ee$
vertices, they have tightly  constrained by the LEP and electroweak precision observables (EWPO). 
The CP-conserving operators in the SILH basis that affect the
top quark interactions at leading order in the $t \bar{t} + \text{jet}$ are listed below:

%------------------------------------------------
\begin{eqnarray}\label{opt}
\mathcal{O}_{uW} &=&{\bar Q}_L\sigma^i H^c
\sigma^{\mu\nu} u_R\,
W_{\mu\nu}^i , \nonumber\\
\mathcal{O}_{uB} &=& {\bar Q}_L H^{c}
\sigma^{\mu\nu}u_R\,
B_{\mu\nu},  \nonumber\\
\mathcal{O}_{uG} &=& 
{\bar Q}_{L} H^{c}
\sigma^{\mu\nu}
\lambda^a u_R \,
G_{\mu \nu}^a,\, \nonumber \\
\mathcal{O}_{HQ} &=&
\left(\bar
Q_{L}\gamma^\mu Q_{L}\right)
\big(H^\dagger
{\overleftrightarrow D}_{\mu}
H \big) \nonumber \\
\mathcal{O}'_{HQ} &=& \left(\bar Q_{L}
\gamma^\mu \sigma^{i} Q_L\right) 
\big(H^\dagger\sigma^{i}
{\overleftrightarrow D}_{\mu}
H \big) , \nonumber \\
\mathcal{O}_{Hu} &=& 
\left(\bar u_{R} \gamma^\mu
u_{R} \right)
\big( H^{\dagger}
{\overleftrightarrow D}_{\mu}
H \big)   ,
\end{eqnarray}
%------------------------------------------------

where  the left-handed and right-handed quarks are denoted by 
$Q_{L}$ and $u_{R}$, respectively and $ H^{\dagger}
{\overleftrightarrow D}_{\mu}H \equiv H^{\dagger} D_{\mu} H - D_{\mu} H^{\dagger}H$.

Among the mentioned operators,  $\mathcal{O}_{uG}$ modifies
the interaction of the top quark and gluons, {\it i.e.} $gt\bar{t}$ and generates 
the new four-leg interaction of $hgt\bar{t}$ which contributes to the $t\bar{t}g$ production.
The $\mathcal{O}_{uW}$, $\mathcal{O}_{uB}$,  $\mathcal{O}_{HQ}$, $\mathcal{O}'_{HQ}$, and 
$\mathcal{O}_{Hu}$ operators modify the interactions between the top quark, photon and the $Z$ boson.

The $\mathcal{O}_{uW}$ and $\mathcal{O}_{uB}$ operators modify
the oblique parameters $S,T,$ and $U$ at one loop level.  In particular, the $\bar{c}_{uW}$ and $\bar{c}_{uB}$
Wilson coefficients have been constrained at percent level  using the oblique parameters \cite{Englert:2017dev}.
Recent measurements of the $t\bar{t}Z$ and $t\bar{t}W$ processes by the CMS collaboration have provided 
the following bounds on $\bar{c}_{uG}$, $\bar{c}_{uW}$, $\bar{c}_{uB}$, and $\bar{c}_{Hu}$ \cite{cms1,cms2}:
\begin{eqnarray}
-0.14 \leq \bar{c}_{uW} \leq 0.14 ~,~ 0.0 \leq \bar{c}_{uB} \leq 0.13 ~,~ -0.07 \leq \bar{c}_{uG} \leq 0.2~,~ -0.64 \leq \bar{c}_{Hu} \leq  0.12,
\end{eqnarray}
Based on the global fit of the experimental data from the Tevatron, and LHC Runs I and II to the SM effective field theory, 
more stringent bounds on these Wilson coefficients could be derived \cite{r2, gf1, gf2,gf3}.
The derived constraints on the considered Wilson coefficients in this work
from a global fit to the top quark experimental data are \cite{r2}:
\begin{eqnarray}
  -8.2\times 10^{-4} \leq  \bar{c}_{uG} \leq 1.8\times 10^{-3}   ,  -4.6\times 10^{-2} \leq  \bar{c}_{uB} \leq 7.0\times 10^{-2}, -0.593 \leq \bar{c}_{Hu} \leq  0.496, \nonumber \\
   -8.9\times 10^{-3} \leq  \bar{c}_{uW} \leq 6.5\times 10^{-3}, 
  -0.369 \leq  \bar{c}_{HQ} \leq 0.375 ,  -3.92\times 10^{-2}  \leq \bar{c'}_{HQ} \leq 2.27 \times 10^{-2},
\end{eqnarray}

The imaginary parts of some of the coefficients of these operators can be constrained using 
the upper limit on the neutron electric dipole moment. The derived upper bound on  Im($\bar{c}_{uG}$)  at $95\%$ CL 
is of the order of $10^{-4}$ \cite{Contino:2013kra}. 

In order to calculate the impacts of the operators on the
top quark pair production in association with a jet,  {\tt MadGraph5\_aMC@NLO}~\cite{Alwall:2011uj,Alwall:2014bza,Alwall:2014hca}
package is used. The effective SM Lagrangian introduced in Eq.~\eqref{eq:SMEFT-2} is implemented in the FeynRule 
 program~\cite{Alloul:2013bka} and then the  Universal FeynRules Output (UFO) model~\cite{Degrande:2011ua} 
 is fed to the {\tt MadGraph5\_aMC@NLO} program. 
 Top quark pair is produced with up to one additional parton in the final state using leading-order matrix elements. 
 The 0-, 1-parton events are merged using the MLM matching scheme \cite{mlm}.

Figure \ref{fig:sigma_couplings} shows the  $t \bar{t} + $(jet)  production cross section
as a function of the centre-of-mass energy at leading order for three signal scenarios as well as the SM background.   
In this figure the Wilson coefficients are normalised to the {\it bar} notation,  $\bar{c}_{X} = c_{X}v^{2}/\Lambda^{2}$, and the
$\mathcal{O}_{uG}$, $\mathcal{O}_{uW}$, and $\mathcal{O}_{uB}$ operators are individually switched on.
As it can be seen, there is a significant enhancement which occurs at
top quark pair threshold. For the SM, the production rate
approximately falls down as $1/\sqrt{s}$.
At $\sqrt{s} =  3$ TeV, the cross section due to the presence of operator $\mathcal{O}_{uG}$ with $\bar c_{uG} = 0.03$ increase 
by a factor of around two with respect to the SM 
while the enhancements arising from $\mathcal{O}_{uW}$ and $\mathcal{O}_{uB}$ with $\bar c_{uW} = 0.03$ 
and $\bar c_{uB} = 0.03$  are at the order of $20$ and $40$, respectively.
Such raises of the cross section occur because of the momentum dependence 
of the $\mathcal{O}_{uW}$, $\mathcal{O}_{uB}$, and $\mathcal{O}_{uG}$ operators.
The $\mathcal{O}_{uW}$ and $\mathcal{O}_{uB}$ operators lead to much larger increase in the cross section of
signal with respect to $\mathcal{O}_{uG}$
because the involved virtual photon and $Z$ boson momenta could grow up to
 the total electron-positron center-of-mass energy while 
less momentum is running to the $\mathcal{O}_{uG}$ vertex. 
As mentioned previously, in this analysis the main aim is to 
examine the potential of the future lepton colliders to probe the SMEFT 
via the top quark pair production in association with a jet. 
In this work, in addition to  $\bar{c}_{uB}$, $\bar{c}_{uG}$ and $\bar{c}_{uW}$, we
examine $\bar{c}_{Hu}$, $\bar{c}_{HQ}$, and $\bar{c}'_{HQ}$ . 

%------------------------------------------------
\begin{figure*}[htb]
\vspace{0.50cm}
\begin{center}
\resizebox{0.55\textwidth}{!}{\includegraphics{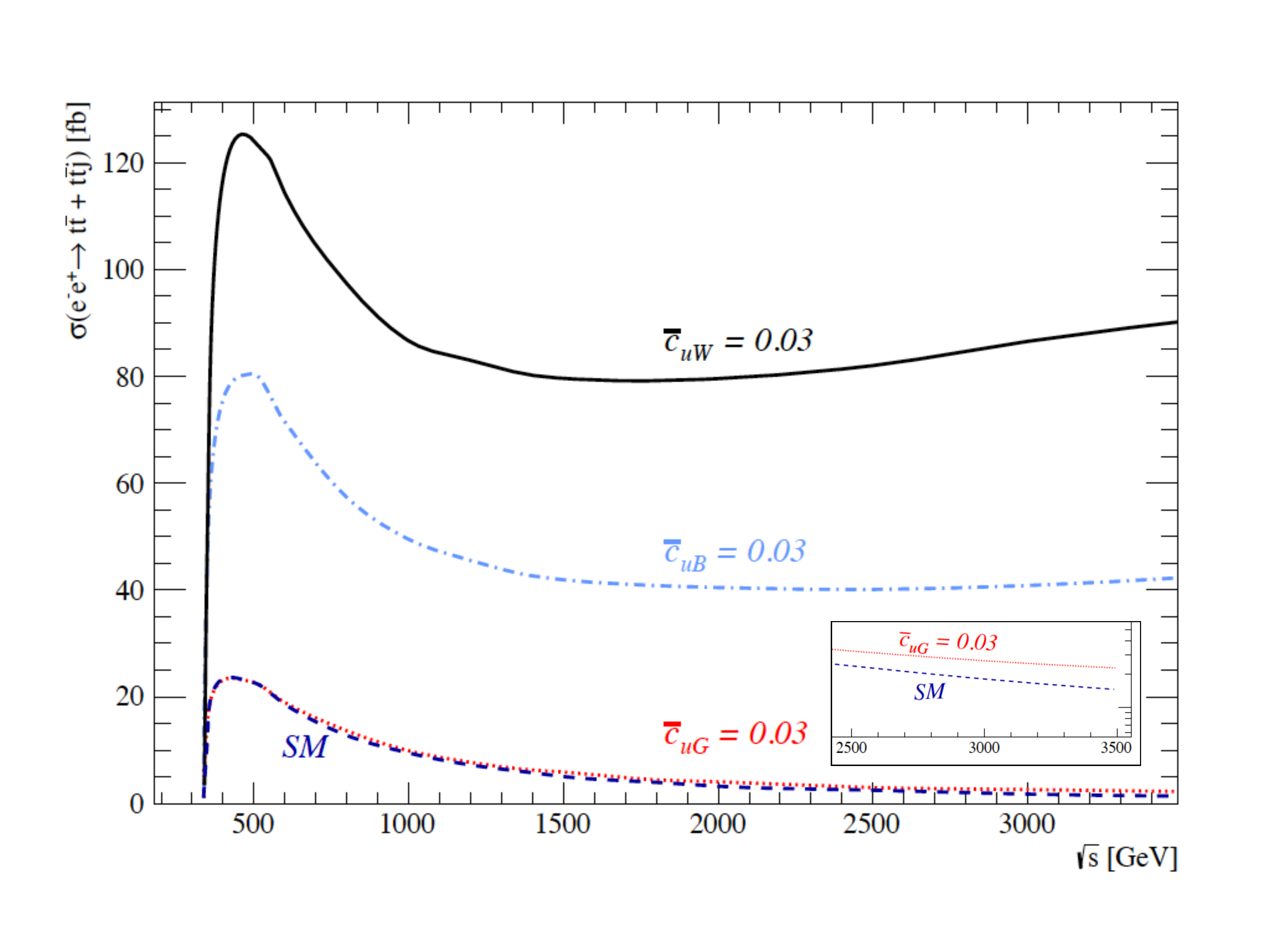}}
\caption{{\small The leading order cross section for the production of $e^{-} e^{+} \rightarrow t \bar{t} + t \bar{t}j$ (merged using MLM) versus  
the center-of-mass energy. The results are shown for the SM and for the signal scenarios  in bar notation $\bar{c}_{X} = c_{X}v^{2}/\Lambda^{2}$
with the assumptions
of $\bar{c}_{uG} = 0.03$, $\bar{c}_{uW} =0.03$, and $\bar{c}_{uB} =0.03$. The cross sections have been calculated with a minimum cut  of $p_{\rm T} \geq 20$ GeV
 on the gluon. The small plot in the bottom shows the cross section for the SM and for $\bar{c}_{uG} = 0.03$ in log-scale. } \label{fig:sigma_couplings}}
\end{center}
\end{figure*}
%------------------------------------------------

%------------------------------------------------
\begin{figure*}[htb]
	\vspace{0.50cm}	
		\begin{center}
	\includegraphics[width=0.46\textwidth]{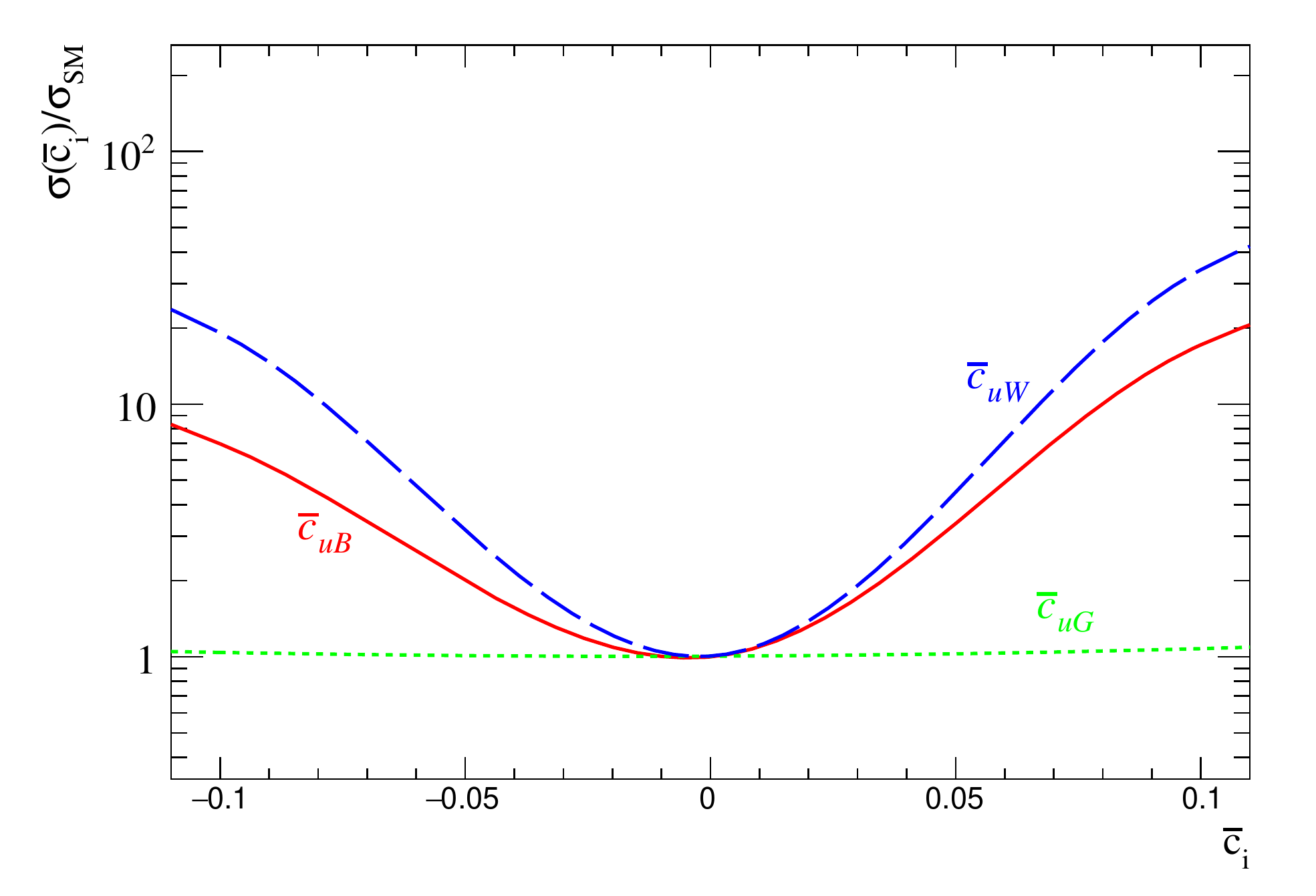}
		\includegraphics[width=0.47\textwidth]{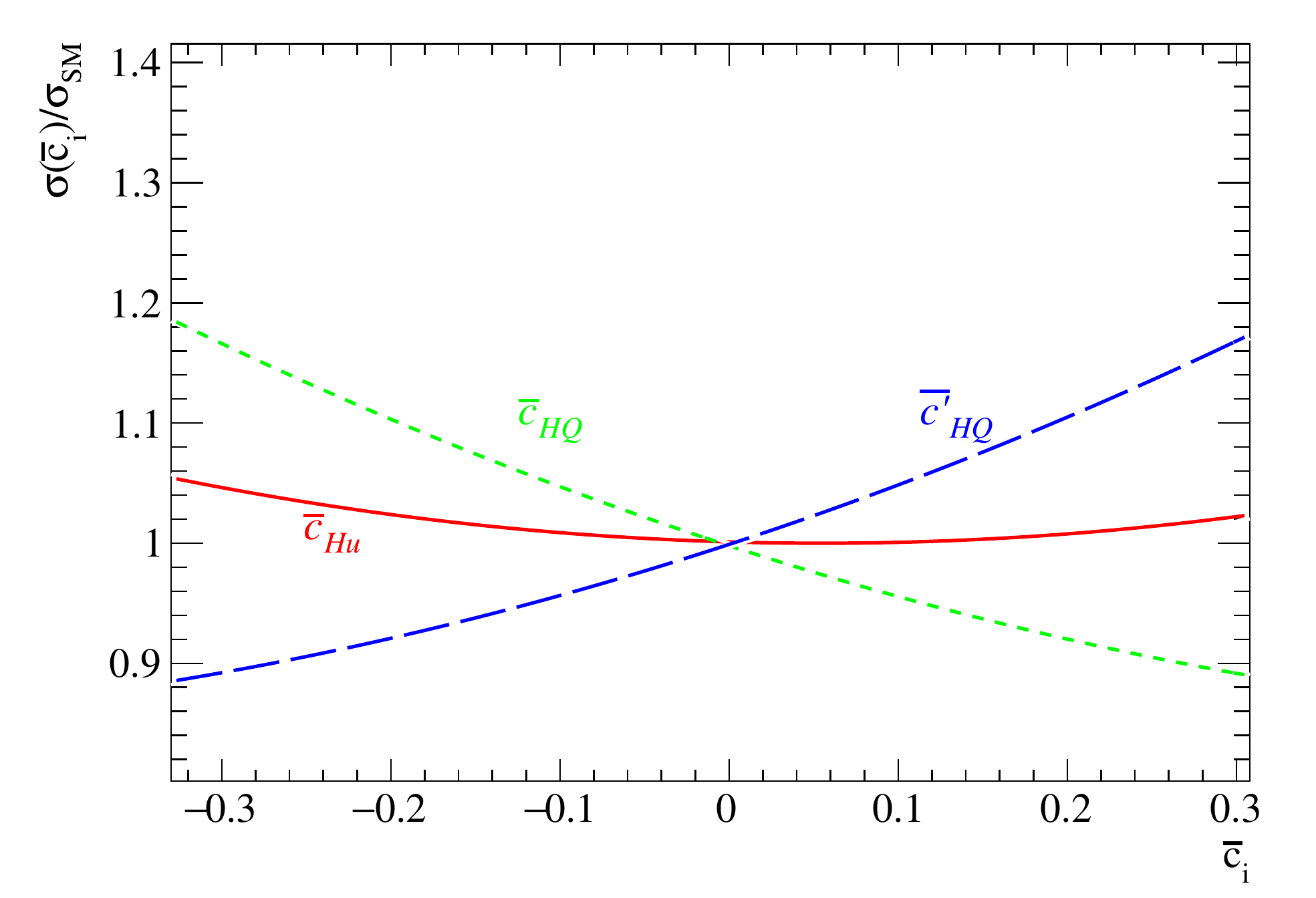}
		\caption{{\small  Ratio of $t\bar{t}+$jet production  cross section in the SMEFT to the SM 
		 in electron-positron collisions versus the Wilson coefficients. The rates are calculated at  leading-order at $\sqrt{s} = 500$ GeV for the $\bar{c}_{uG}$, $\bar{c}_{uW}$, 
		 $\bar{c}_{uB}$, $\bar{c}_{HQ}$, $\bar{c}'_{HQ}$, and $\bar{c}_{Hu}$.  A minimum cut of 20 GeV has been applied on the  transverse momentum of additional jet.
		 } \label{ratio}}
	\end{center}
\end{figure*}
%------------------------------------------------

Figure~\ref{ratio}  shows  the ratio of the production  cross  section  of signal process in the SMEFT to the SM in  terms  of  the  Wilson  coefficients 
  $\bar{c}_{uG}$, $\bar{c}_{uW}$,  $\bar{c}_{uB}$, $\bar{c}_{HQ}$, $\bar{c}'_{HQ}$, and $\bar{c}_{Hu}$.  
To calculate the cross sections, a minimum cut of 20 GeV has been applied on the gluon $p_{T}$. 
 As it can be seen in the left plot of Fig.\ref{ratio}, there is a remarkable sensitivity to $\bar{c}_{uW}$ and  $\bar{c}_{uB}$ while $\bar{c}_{uG}$ has less
 impact  on  the  production  cross section.   
 From the right plot in Fig.~\ref{ratio},   we observe that  $\bar{c}_{HQ}$, $\bar{c}'_{HQ}$, and $\bar{c}_{Hu}$ 
  have no considerable effect in the cross section with respect to $\bar{c}_{uW}$,  $\bar{c}_{uB}$, and $\bar{c}_{uG}$.
 In the analysis we probe
 these Wilson coefficients at two different center-of-mass energies of
 500 and 3000 GeV.

}
%
%%%%%%%%%%%%%%%%%%%%%%%%%%%%%%%%%%%%%%%%%%        %%%%%%%%%%%%%%%%%%%%%%%%%%%%%%%%%%%%%%%%%%%%%%%%%
\section{ Simulation and details of the analysis }\label{sec:Simulation}
{

In this section, the details of  the simulation and the analysis strategy to examine 
the dimension-six operators mentioned in Eq.\ref{opt} using 
 the  top quark pair production in association with a jet are discussed.  
 Based on the decay of  $t\bar{t}$ system, there are three different final states for the signal:
 fully hadronic, semileptonic,  and dileptonic final states with the branching fractions of $46.2\%$, $43.5\%$, and $10.3\%$, respectively.  
In this work, in order to have a clean signature, we focus on the dileptonic decay channel therefore the final state consists of 
at least two jets from which two are $b$-jets originating from the top
quarks decay,  two opposite sign charged leptons, 
and missing transverse momentum. 
%In this work, for comparison
%the results will be presented for two selection scenarios based on the number of jets.
%In the first scenario (Scenario I),  exactly two b-jets are required and any events with additional jet
%is excluded. In the second scenario (Scenario II), events are allowed to have at least two jets ($n_{\text{jet}} \geq 2$)
%from which exactly two have to be b-jets.

The dominant background processes to the signal considered in the analysis are as follows:
\begin{itemize}
\item{SM production of $t\bar{t}+$jet (merged $t\bar{t}$ and $t\bar{t}+$jet using MLM prescription).}
\item{Single top production $tWj$.}
\item{$e^{-}e^{+} \rightarrow Z^{*}Z^{*}V^{*} \rightarrow 2\ell$ + jets+missing momentum, where $V = \gamma,Z$.}
\item{$e^{-}e^{+} \rightarrow W^{*}W^{*}V^{*} \rightarrow 2\ell$ + jets+missing momentum, where $V = \gamma,Z$.}
\item{$e^{-}e^{+} \rightarrow V^{*}V^{*}V'^{*}V'^{*} \rightarrow 2\ell$ + jets+missing momentum, where $V,V' = W^{\pm},Z,\gamma$.}
\end{itemize}

where $\ell = e, \mu$. The SM background processes and signal events are generated using
 the {\tt MadGraph5\_aMC@NLO}~\cite{Alwall:2011uj,Alwall:2014bza,Alwall:2014hca} event generator. 
The  $t\bar{t}+$jet sample is a merged $t\bar{t}$ and $t\bar{t}+$ jet sample using the  MLM merging prescription \cite{mlm}. 
In  merging process, the xqcut variable defined as the minimal distance between partons at {\tt MadGraph} level
and  qcut  variable which is  the matching scale in  {\tt  PYTHIA} ~\cite{Sjostrand:2014zea,Sjostrand:2007gs}  
are set to 20 GeV and 30 GeV, respectively. 
These choices for xqcut and qcut lead to a smooth transition between events
 with 0 and 1 jet in the differential jet rate distribution.
 In the event generation process, the SM input parameters are considered as~\cite{Tanabashi:2018oca}:
 
%--------------------------------
\begin{eqnarray}
m_t &=& 173.34 \, {\text {GeV}} \, \, {\text { for the top quark mass, }} \nonumber \\
m_W &=& 80.385 \, {\text {GeV}} \, \, {\text {for the $W$ boson mass, }}  \nonumber \\
m_Z &=& 91.187 \, {\text {GeV}} \, \, {\text {for the $Z$ boson. }}
\end{eqnarray}
%--------------------------------

The generated samples are passed through the {\tt PYTHIA 6}~\cite{Sjostrand:2014zea,Sjostrand:2007gs} 
for parton shower, hadronization, and decay of unstable particles.
In order to take into account detector effects, 
we use the {\tt Delphes 3.4.1}~\cite{deFavereau:2013fsa} 
by which an ILD-like detector~\cite{Behnke:2013lya} is simulated.
For jet reconstruction, the anti-$k_{t}$ algorithm~\cite{Cacciari:2008gp} 
based on the FastJet package~\cite{Cacciari:2011ma} with the cone size parameter $R=0.5$ is employed.
The $b$-tagging efficiency and misidentification rates are applied depending on the jet transverse momentum~\cite{Behnke:2013lya} .
The efficiency of  $b$-tagging  for a jet with $p_{\rm T} = 40$ GeV is  $60\%$,
and the charm-jet and light flavour jets misidentification rates are $14\%$ and $1.1\%$, respectively.

To select signal events, it is required to have exactly two  same flavour opposite sign isolated charged leptons (either electron or muon) 
 with the transverse momentum $p_T \geq 20$ GeV and the pseudorapidity $|\eta| \leq 2.5$. 
Each event is required to have at least two jets from which only two must be b-tagged.
Jets are required to have  $p_{\rm T} \geq 20$ GeV and $|\eta| \leq 2.5$.
In order to make sure all selected objects are well isolated,
we require that the angular separation $\Delta R_{i, j} = \sqrt{ (\Delta \phi)^2+(\Delta \eta)^2} \geq 0.4$, where $i,j = \ell$ and jets.
The magnitude of missing transverse momentum is required to be larger than $20$ GeV.

In order to suppress the contributions of the SM background, a multivariate technique is utilised 
\cite{Hocker:2007ht,Stelzer:2008zz,Therhaag:2009dp,Speckmayer:2010zz,Therhaag:2010zz}.
Particular, in this work the gradient Boosted Decision Trees (BDTG) is used for separating the signal from backgrounds
and to achieve the best sensitivity. 
After the cuts described previously (preselection cuts)  the cross section of signal and the background processes  for the center-of-mass energy of 3000 GeV
are presented in Table \ref{Table:Cut-Table-3TeV}. The signal cross section is presented for three different
 scenarios of $\bar{c}_{uW} = \bar{c}_{uB} = 0.1$, $\bar{c}_{uW} = \bar{c}_{uG} = 0.1$, 
$\bar{c}_{uB} = \bar{c}_{uG} = 0.1$.
The applied cuts are in general loose on a single variable and are not able to suppress a
 considerable fraction of  background events while reducing  the signal events. 
 Therefore, a gradient BDT is trained to achieve a better discrimination of
 signal from background processes. All the backgrounds are considered in the 
 training according  to their associated weights. 
For the sake of obtaining an effective separation of signal from 
the background events, an appropriate set of variables needs to be
chosen. In this analysis, the following variables are used: the scalar sum of transverse momentum of
the leptons and jets, $H_{T}$;  invariant mass of the two b-jets ($m_{b_{1}b_{2}}$);  $\eta$ of the leading lepton; 
$\eta$ of the leading and sub-leading b-jets; (v) the angular separation of two b-tagged jets $\Delta R(b_{1},b_{2})$.
In Fig.\ref{fig:TMVA-1}, the distributions of some of  variables are depicted. 
These distributions presented in Fig.\ref{fig:TMVA-1} are corresponding to four
 signal scenarios with $\bar{c}_{uW} = \bar{c}_{uB} = 0.1$, $\bar{c}_{uW} = \bar{c}_{uG} = 0.1$, 
$\bar{c}_{uB} = \bar{c}_{uG} = 0.1$,  and $\bar{c}_{Hu} = \bar{c'}_{HQ} = 0.1$ at the center-of-mass energy of 3000 GeV. 
For all signal scenarios, the same  input variables are used for BDT training. 

%------------------------------------------------
\begin{figure*}[htb]
	\vspace{0.50cm}	
	\begin{center}
		\resizebox{0.48\textwidth}{!}{\includegraphics{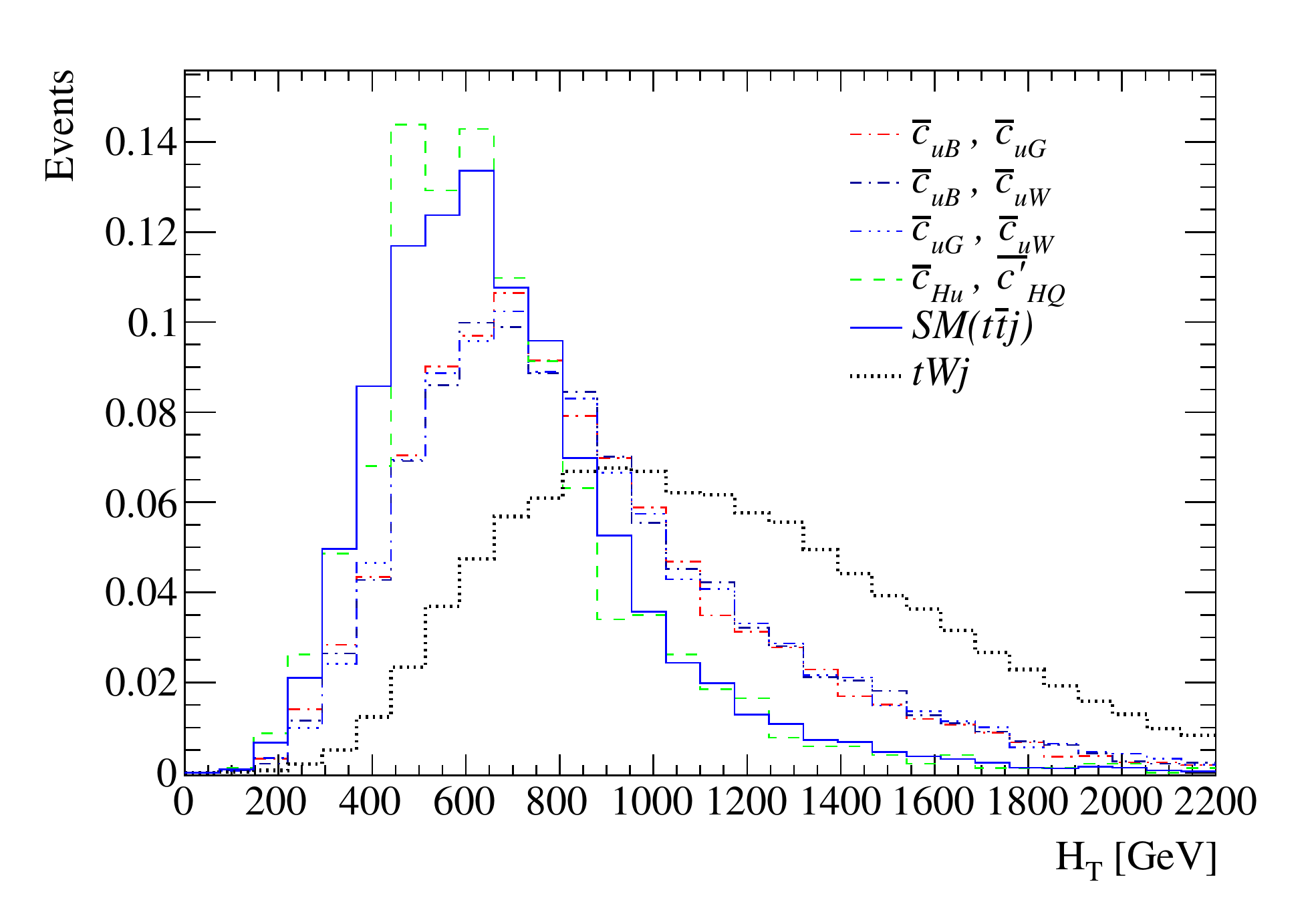}}
		\resizebox{0.48\textwidth}{!}{\includegraphics{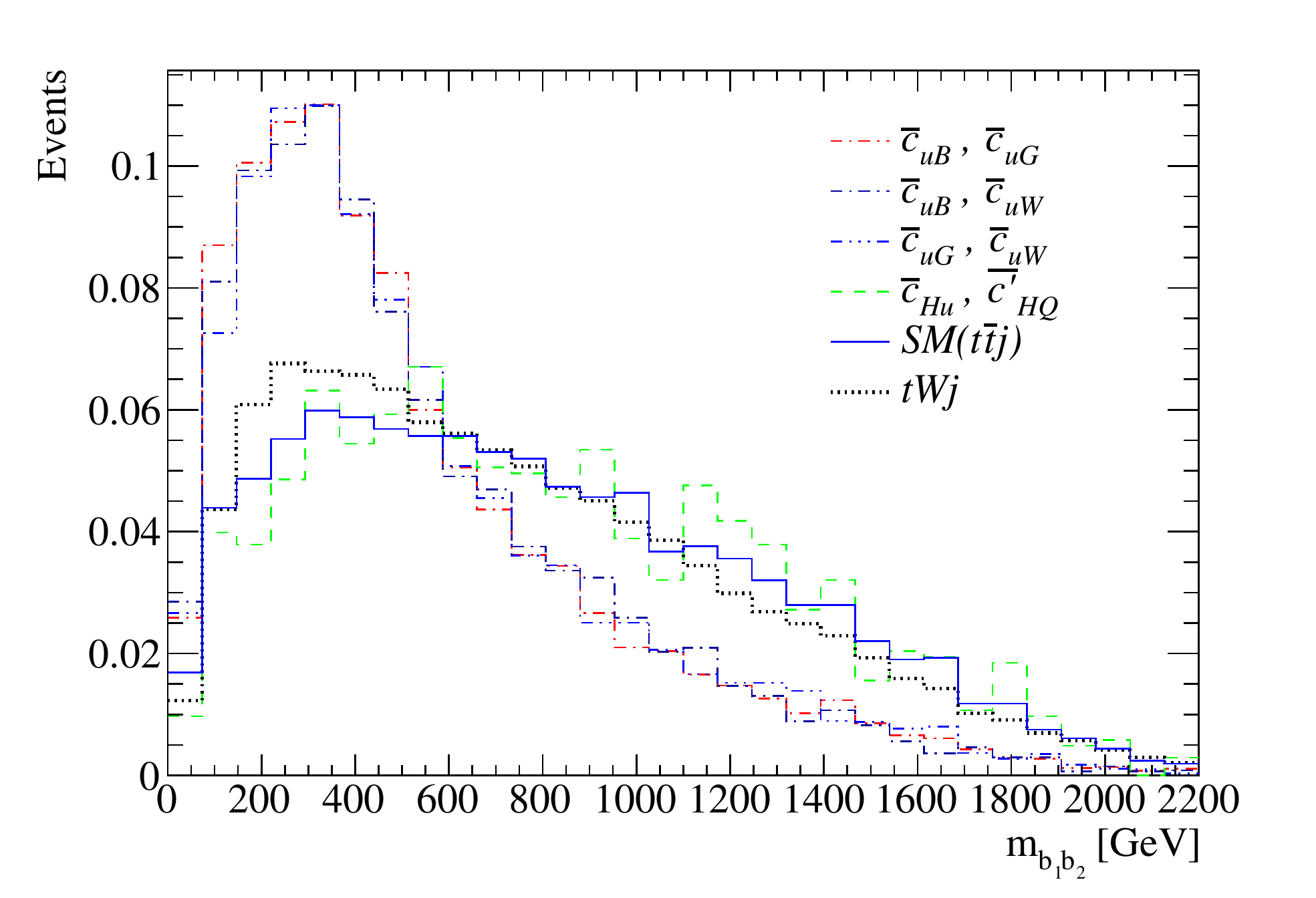}}
		\resizebox{0.48\textwidth}{!}{\includegraphics{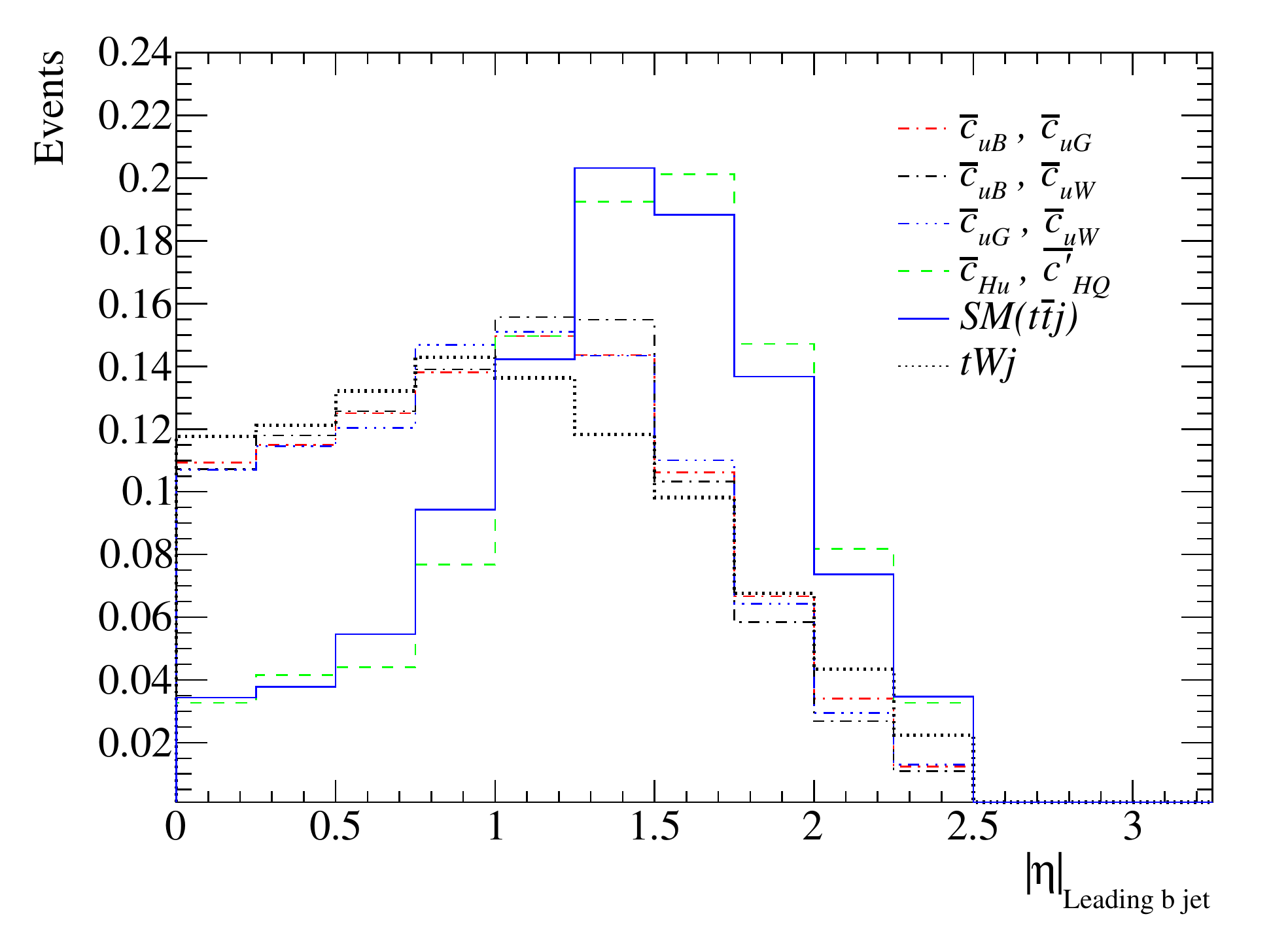}}
		\resizebox{0.46\textwidth}{!}{\includegraphics{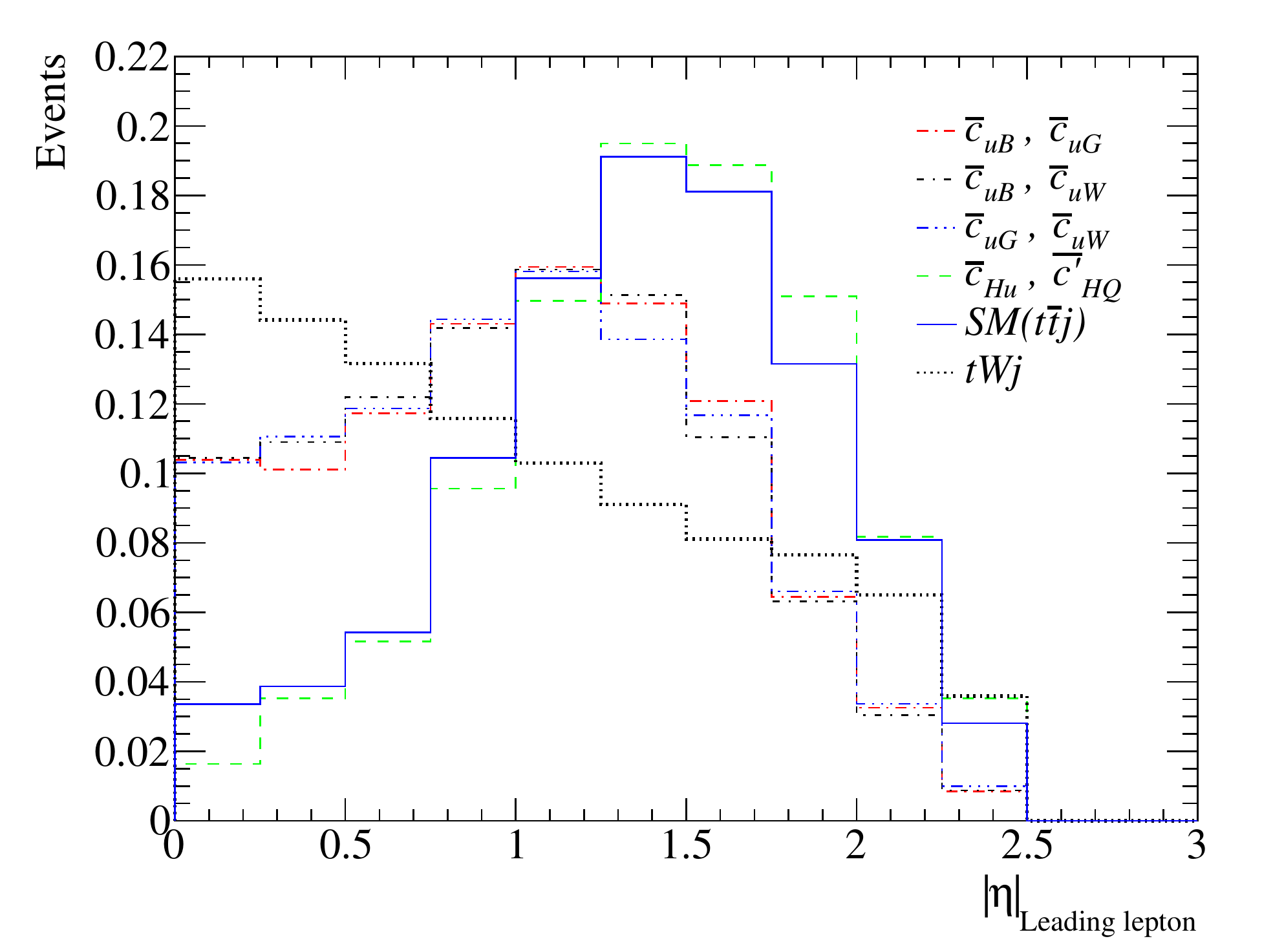}}
		\caption{{\small The normalised distributions of some of the input variables to the multivariate analysis. The plots show distributions of $H_T$ (top left),   invariant mass of b-jets (top right)
		pseudorapidity of  leading b-jet (bottom left) and leading lepton  (bottom right) at $\sqrt{s} = $ 3000 GeV. } \label{fig:TMVA-1}}
	\end{center}
\end{figure*}
%------------------------------------------------

For instance, the BDTG output distribution for the signal scenario of $\bar{c}_{uB} = \bar{c}_{uG} = 0.1$ is shown in Fig.\ref{fig:cubcug_BDTG}. 
Contrary to  the overwhelming contribution of backgrounds, 
the gradient BDT performs well. 
The output of BDTG has been checked
 in terms of the power of discrimination from the receiver operator characteristic (ROC) of the output of BDTG output. 
 The  optimum cut on the BDTG response is chosen so that  the best  sensitivity is achieved.
  %------------------------------------------------
\begin{figure*}[htb]
\vspace{0.50cm}	
\begin{center}
\resizebox{0.450\textwidth}{!}{\includegraphics{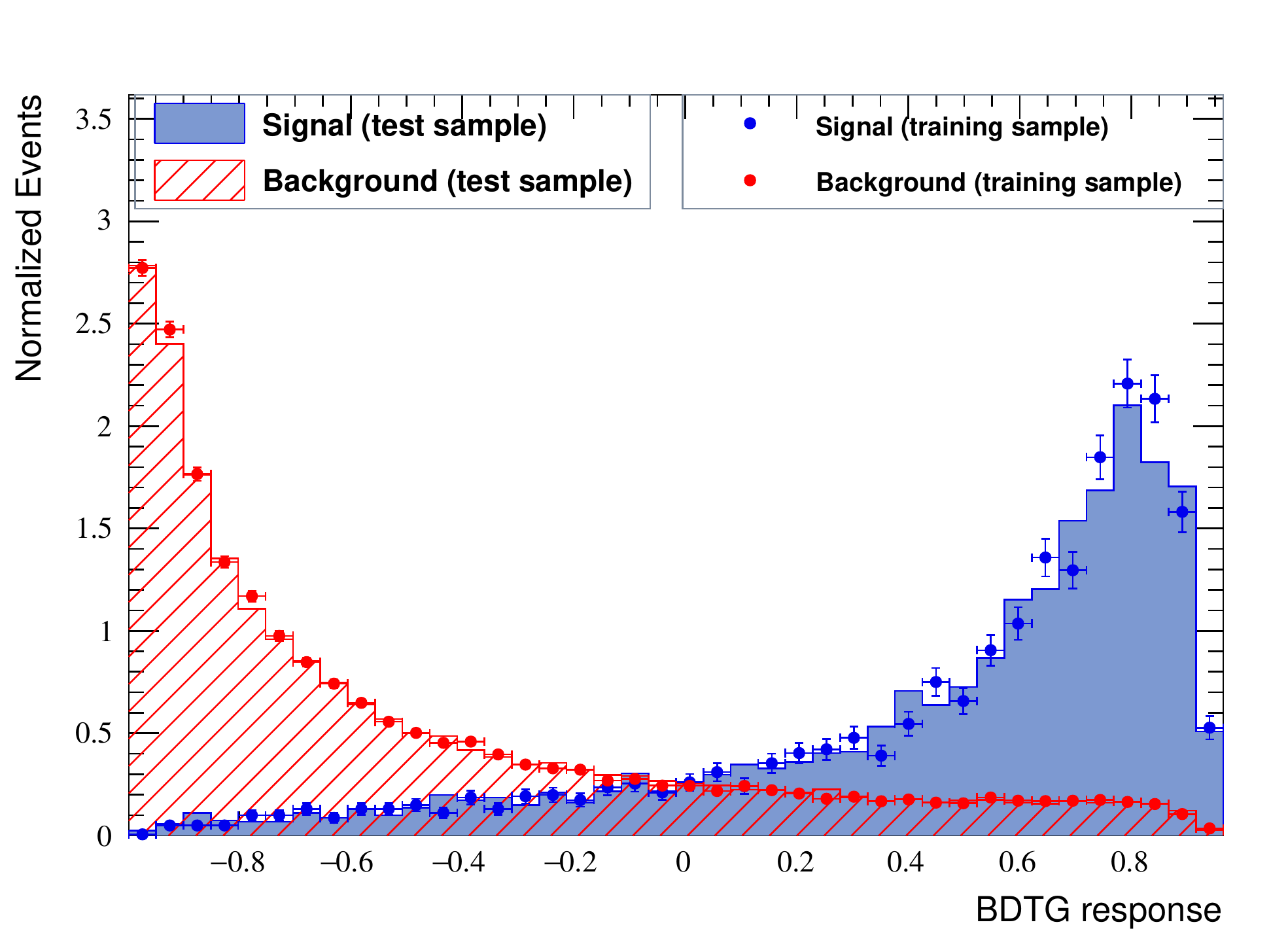}}
\caption{{\small The distribution of the gradient BDT output for the  signal with $c_{uB}= c_{uG} = 0.1$ and for the  SM backgrounds at the
		the center-of-mass energy of 3000 GeV.  } \label{fig:cubcug_BDTG}}
\end{center}
\end{figure*}
%------------------------------------------------

  Separate analyses are performed at the center-of-mass energies of 500 and 3000 GeV.
The signal and background cross sections at $\sqrt{s} = 3000$ GeV after performing the multivariate analysis are given  in Table \ref{Table:Cut-Table-3TeV} for the signal scenario with 
   $ \bar{c}_{uW} = \bar{c}_{uB} = 0.1$.  
 As we can see from  Table \ref{Table:Cut-Table-3TeV}, the main background contributions come  
from the $tWj$ and $t \bar{t}+$jet processes. 

%--------------------------------
\begin{table*}[ht]
	\begin{center}
		\begin{tabular}{ccccccc}\hline
		 $\sqrt{s} = 3000$ GeV &  Couplings   &   Signal  &  $t\bar{t}+$jet &  $tWj$  & $WWV+ZZV$  & $VVV'V'$ \\ \hline
		 MVA  & $(\bar{c}_{uW},\bar{c}_{uB})$ &  4.42  &  0.0021    & 0.0041   &  0.0005 & 0.000043   \\ \hline
		 $\sqrt{s} = 500$ GeV &  Couplings   &   Signal  &  $t\bar{t}+$jet &  $tWj$  & $WWV+ZZV$  & $VVV'V'$ \\ \hline
		 MVA  & $(\bar{c}_{uW},\bar{c}_{uB})$ &  247.5  &   3.6   &   0.17 &  0.03&   0.000023 \\ \hline
		\end{tabular}
	\end{center}
	\caption{Expected cross sections of signal and background processes at $\sqrt{s} = 500$ and 3000 GeV after performing the multivariate analysis. The signal cross section is  presented for 
		$\bar{c}_{uB} = \bar{c}_{uW} =0.1$ in the unit of fb.  }
	\label{Table:Cut-Table-3TeV}
\end{table*}
%--------------------------------

We note that  the considered operators affect the background processes.
In this work, after the  cuts and multivariate analysis,  backgrounds are suppressed remarkably and the impacts of the
included dimension six operators on the backgrounds are not sizeable. 
For instance, the change in the cross section of the $tWj$  background at $\sqrt{s} = 500$ GeV in different scenarios are as follows:
\begin{eqnarray}
\Delta\sigma_{tWj} &=&  \sigma_{tWj}(\bar{c}_{uW} = 0.1,\bar{c}_{uG} = 0.1) - \sigma_{tWj}(0.0,0.0)=  0.632 \nonumber \\ 
\Delta\sigma_{tWj} &=&  \sigma_{tWj}(\bar{c}_{uW} = 0.1,\bar{c}_{uB} = 0.1) - \sigma_{tWj}(0.0,0.0)=  0.637 \nonumber  \\
\Delta\sigma_{tWj} &=&  \sigma_{tWj}(\bar{c}_{uB} = 0.1,\bar{c}_{uG} = 0.1) - \sigma_{tWj}(0.0,0.0)=  3.9\times 10^{-3},
\end{eqnarray}
the numbers are given in the unit of fb. The impact of the other operators on $tWj$ is quite negligible.
The deviations that other background processes receive from the operators are not considerable and
are found to be of the order of $\lesssim 10^{-4,-5}$ fb.
In this analysis, we consider the impact of operators on the $tWj$ background when limits are set on the Wilson coefficients. 
The  results and sensitivity estimation will be presented in the next section.

}
%
%%%%%%%%%%%%%%%%%%%%%%%%%%%%%%%%%%%%%%%%%%        %%%%%%%%%%%%%%%%%%%%%%%%%%%%%%%%%%%%%%%%%%%%%%%%%
\section{ Results and discussions }\label{sec:Results}
{
In this section, we  present  the   sensitivity of the future lepton colliders  
to  the  coefficients  of  dimension-six  operators that could be obtained at the center-of-mass 
energies of 500 and 3000 GeV.
We present the  expected bounds at $95\%$ CL on the individual operators as well as marginalised limits over all contributing operators.
Two-dimensional contours of the
expected constraints at $95\%$ CL  are presented in Fig.\ref{fig:Contours3000} and
Fig.\ref{fig:Contours500} for the center-of-mass energies of 3000 GeV and 500 GeV, respectively.
The results at $\sqrt{s}  =3000$ GeV are presented for two integrated luminosities of 300 and 3000 fb$^{-1}$.

As expected among the Wilson coefficients, 
the highest sensitivity belongs to $\bar{c}_{uW}$ then to $\bar{c}_{uB}$ so that at $\sqrt{s} = 3$ TeV with 3 ab$^{-1}$ of data,
one could probe them down to $10^{-3}$. In order to investigate how far these sensitivities are changed 
with including uncertainties, we also present the contours at $95\%$ CL 
by considering $10\%$ uncertainty on the cross sections
of the background processes and a total $10\%$ uncertainty on the efficiency 
of signal. This would loosen the constraints up to around $15\%$. 

The numerical one dimensional constraints at $95\%$ CL  on the
Wilson coefficients at both center-of-mass energies of 500 and 3000 GeV are
given in Table~\ref{res2}.

%------------------------------------------------
\begin{figure*}[htb]
	\vspace{0.50cm}	
	\begin{center}
		\resizebox{0.23\textwidth}{!}{\includegraphics{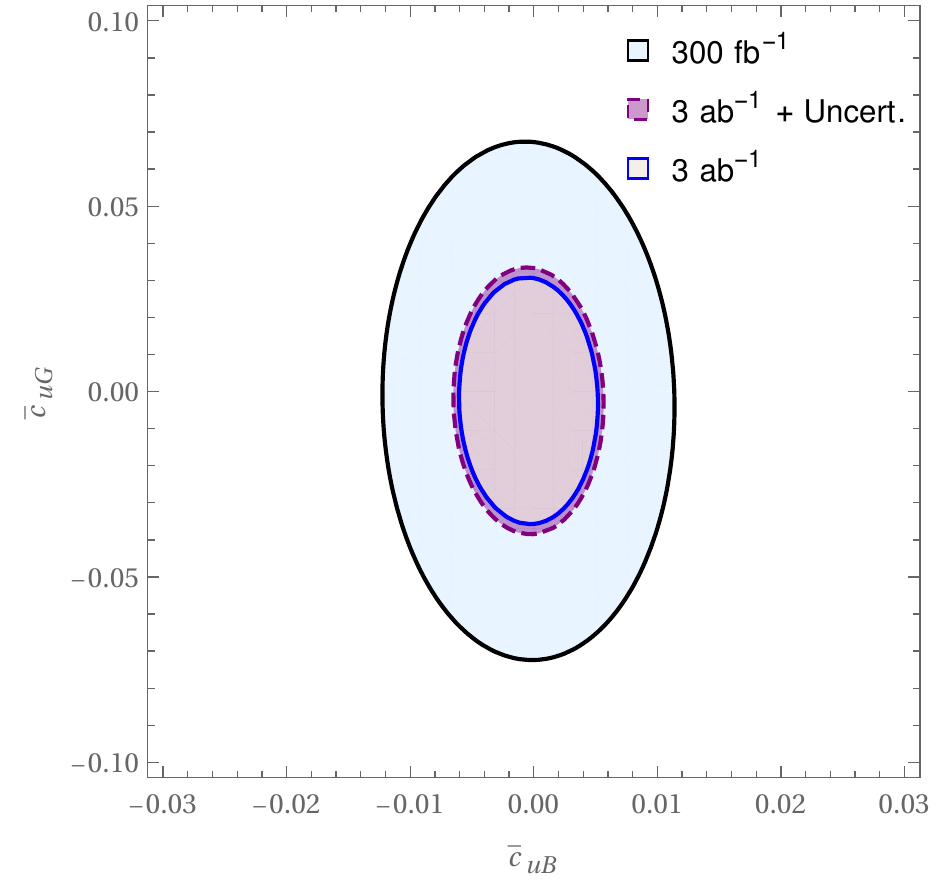}}
		\resizebox{0.23\textwidth}{!}{\includegraphics{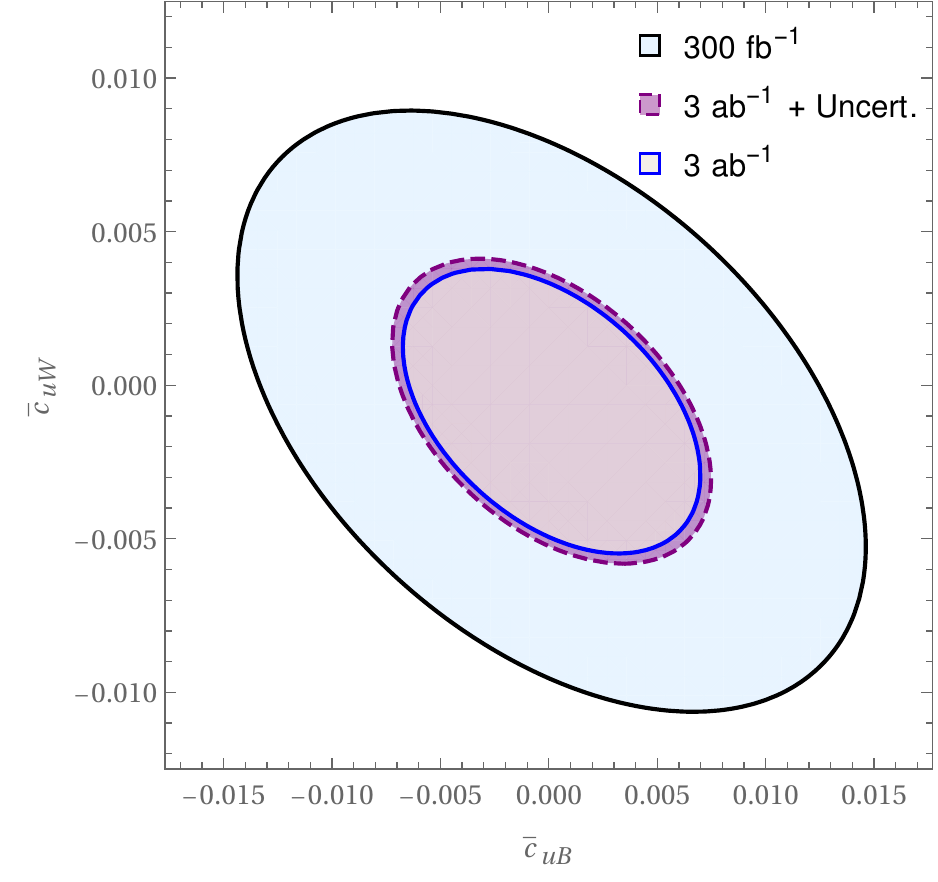}}
		\resizebox{0.23\textwidth}{!}{\includegraphics{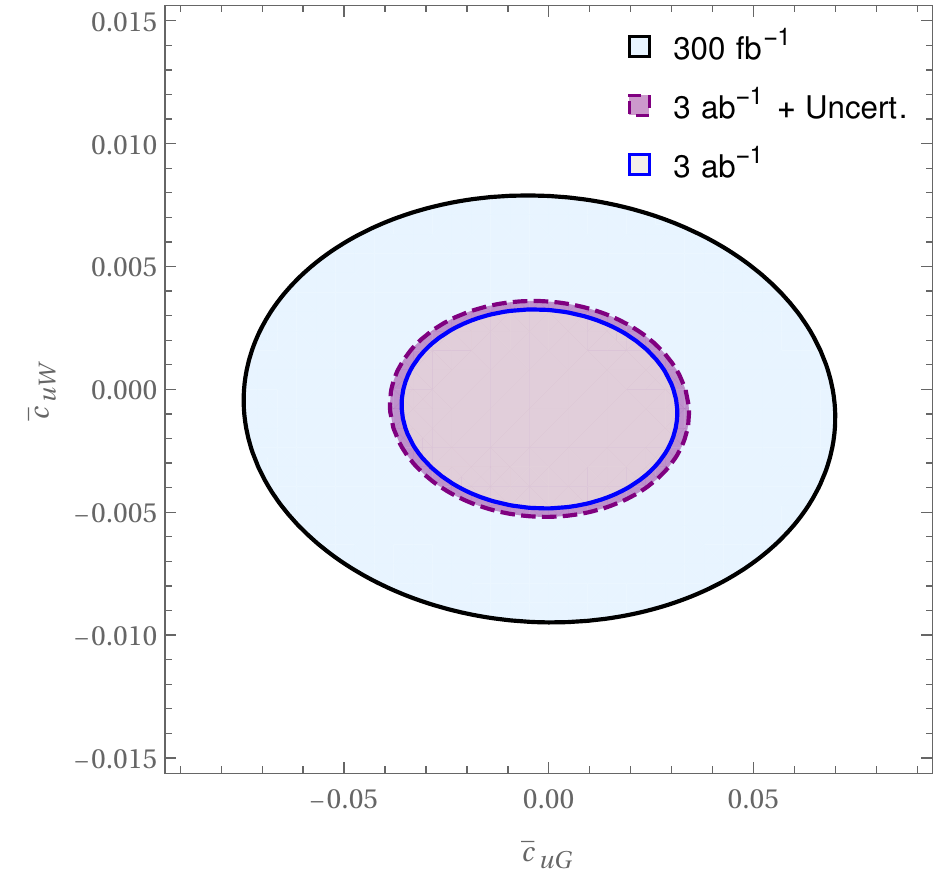}}
		\resizebox{0.22\textwidth}{!}{\includegraphics{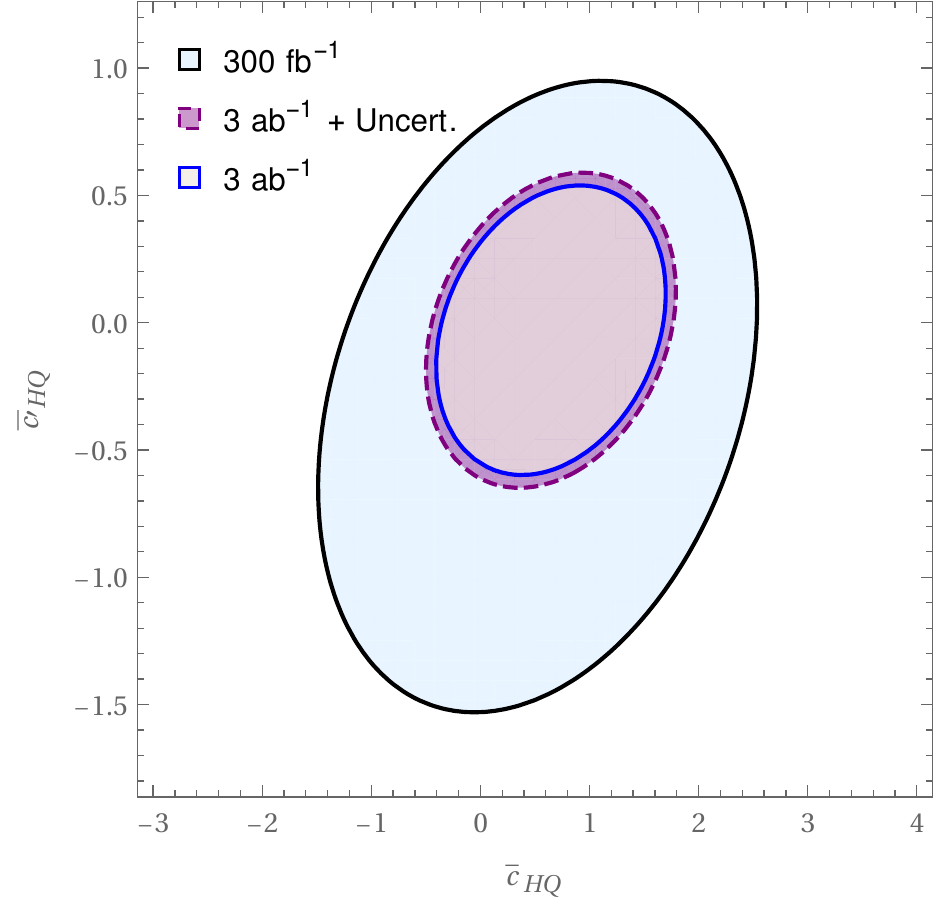}}
		\resizebox{0.23\textwidth}{!}{\includegraphics{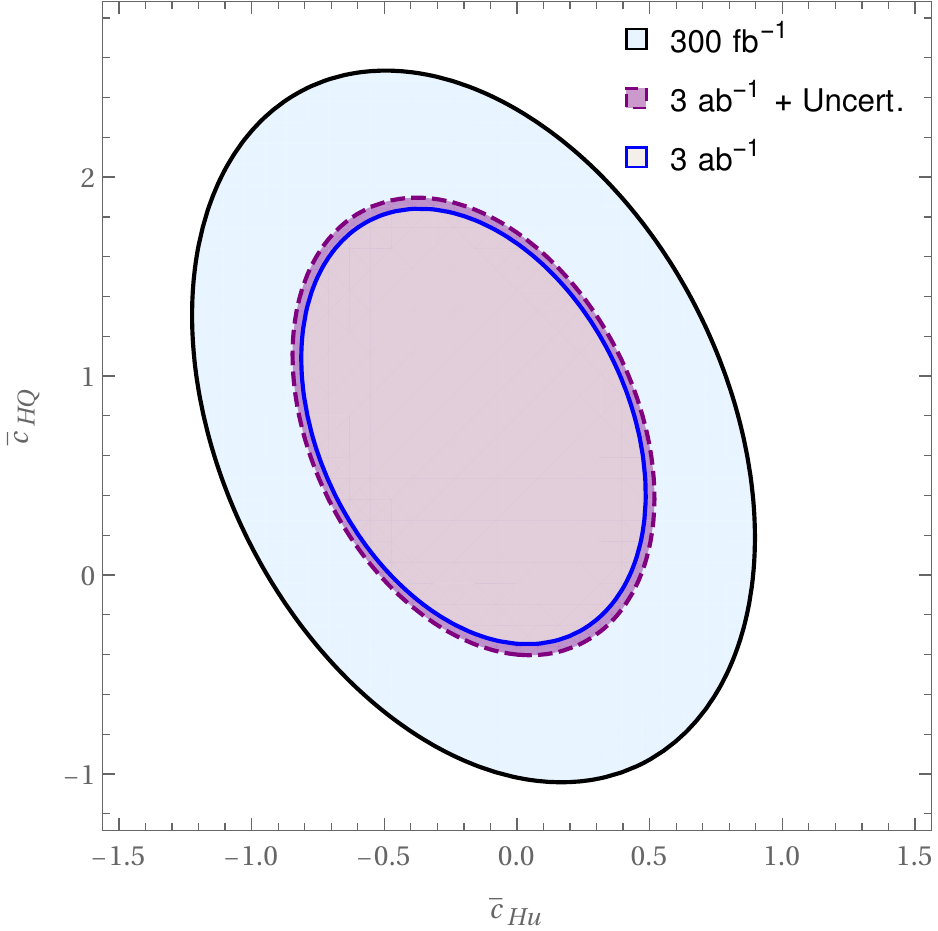}}
	\resizebox{0.23\textwidth}{!}{\includegraphics{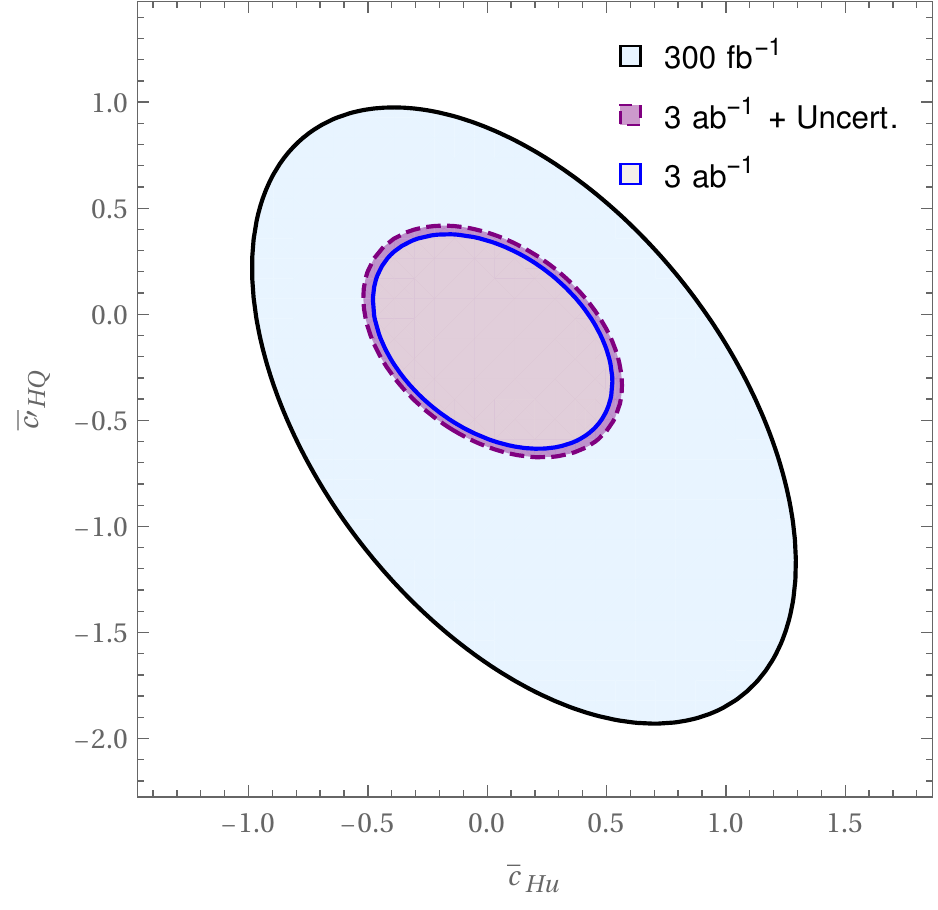}}
	\resizebox{0.23\textwidth}{!}{\includegraphics{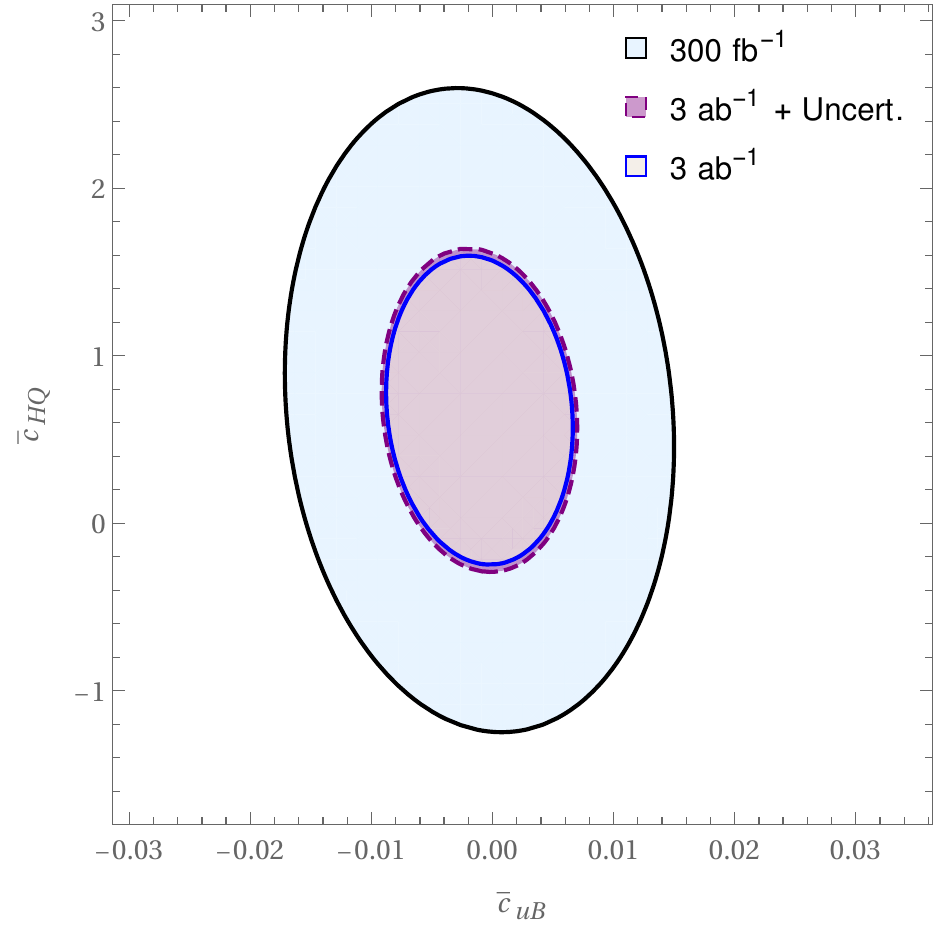}}
	\resizebox{0.23\textwidth}{!}{\includegraphics{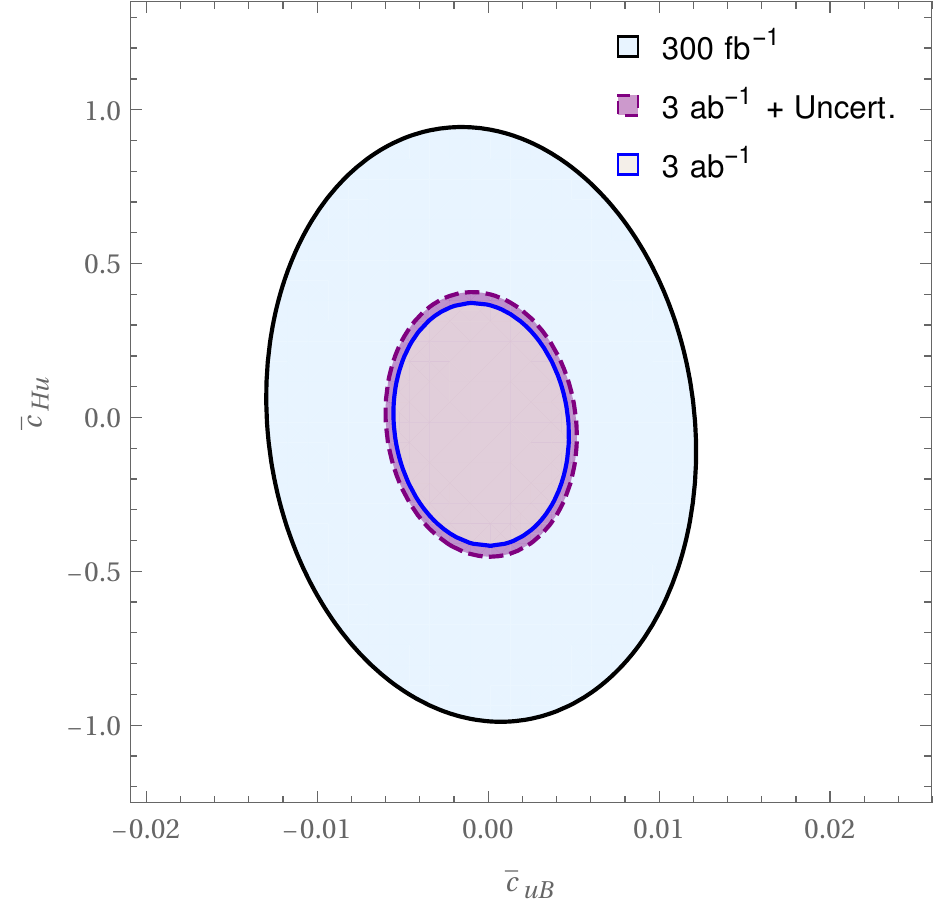}}
	\resizebox{0.23\textwidth}{!}{\includegraphics{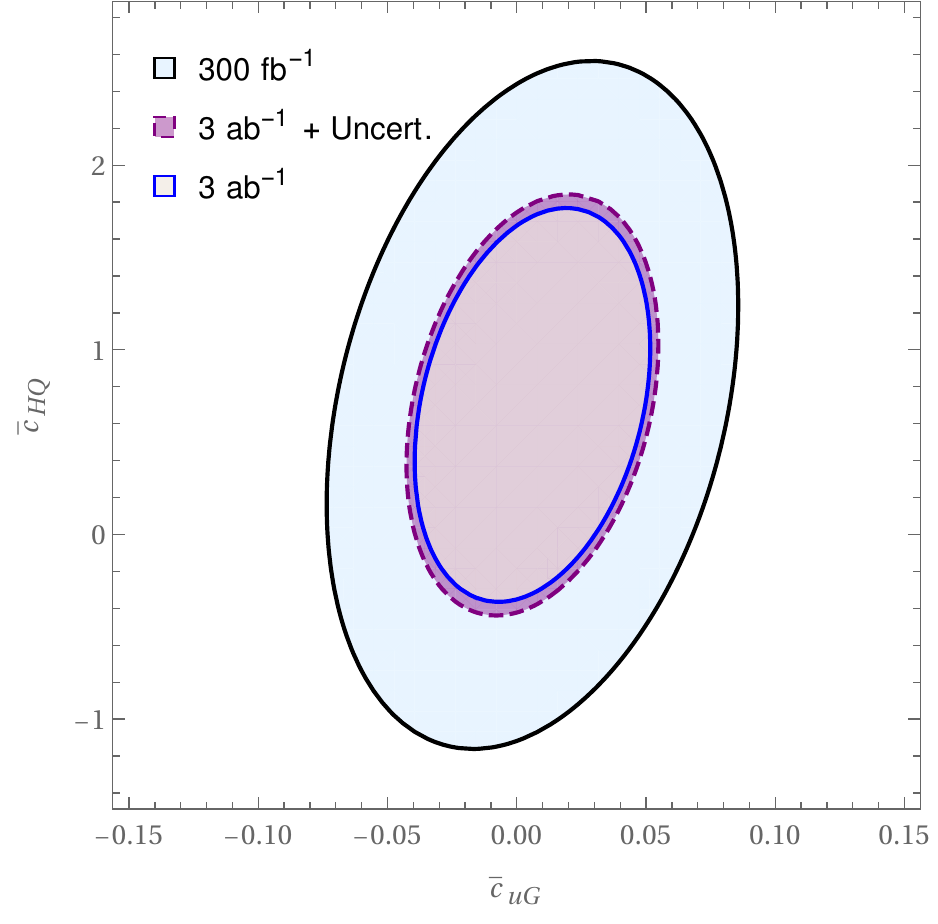}}
	\resizebox{0.23\textwidth}{!}{\includegraphics{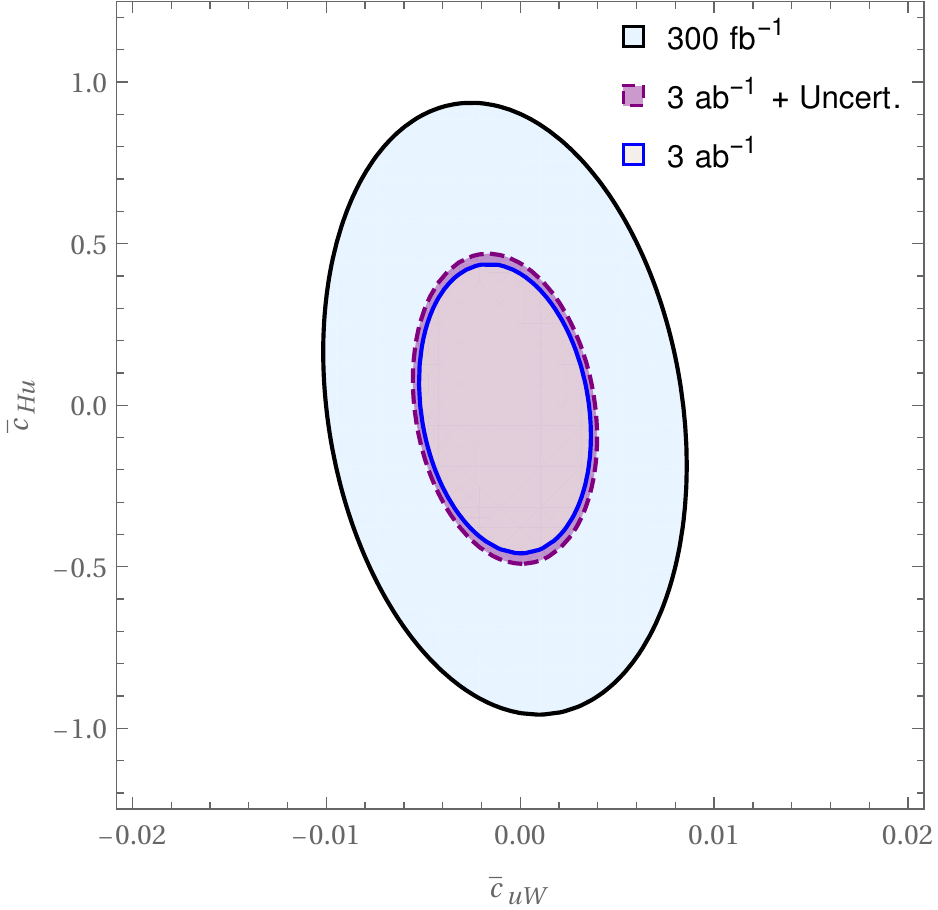}}
	\resizebox{0.23\textwidth}{!}{\includegraphics{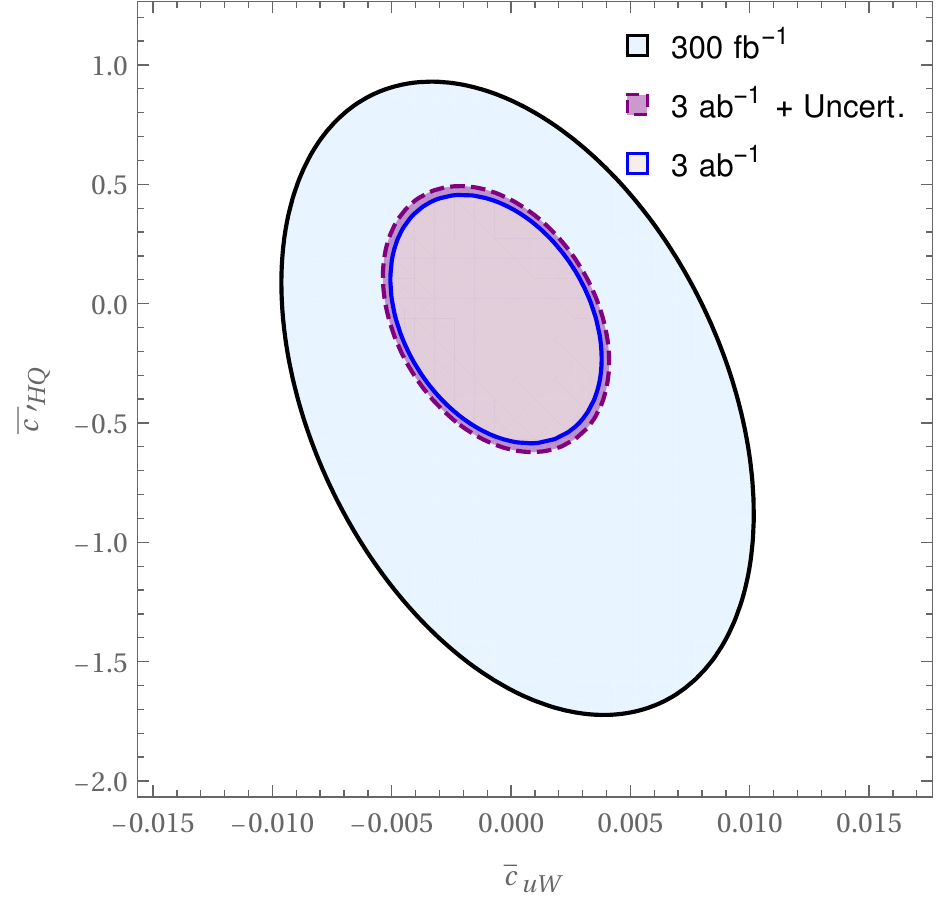}}
	\resizebox{0.23\textwidth}{!}{\includegraphics{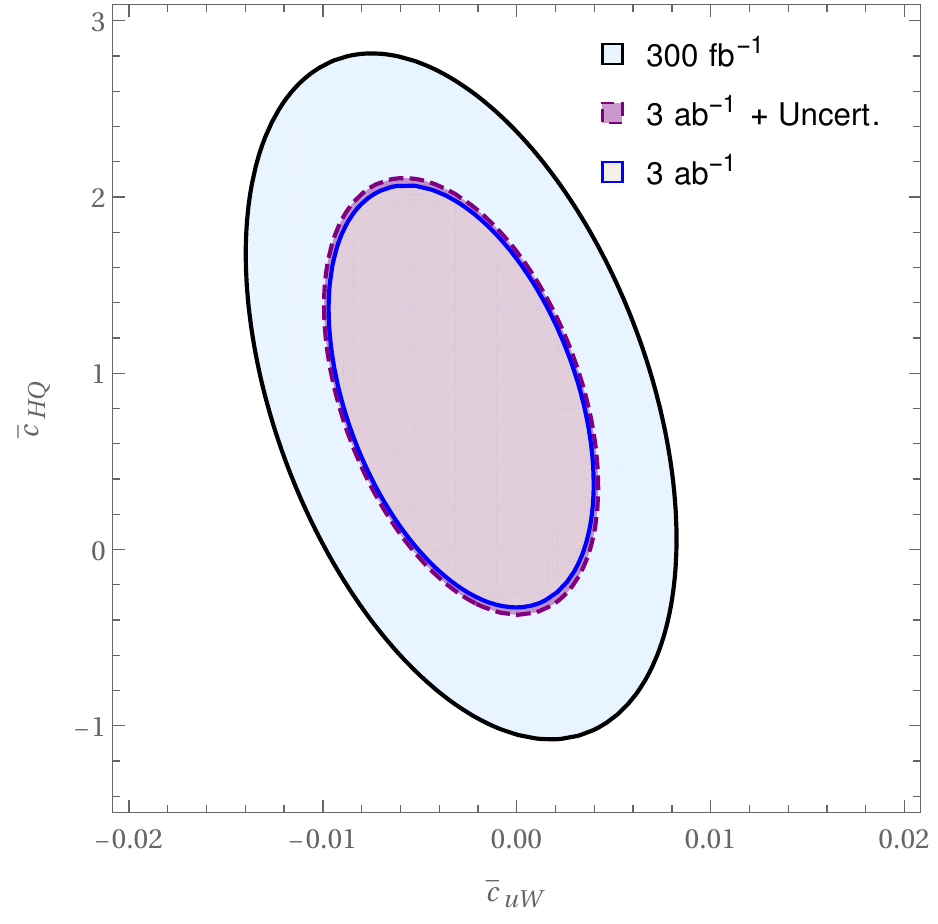}}
		\resizebox{0.23\textwidth}{!}{\includegraphics{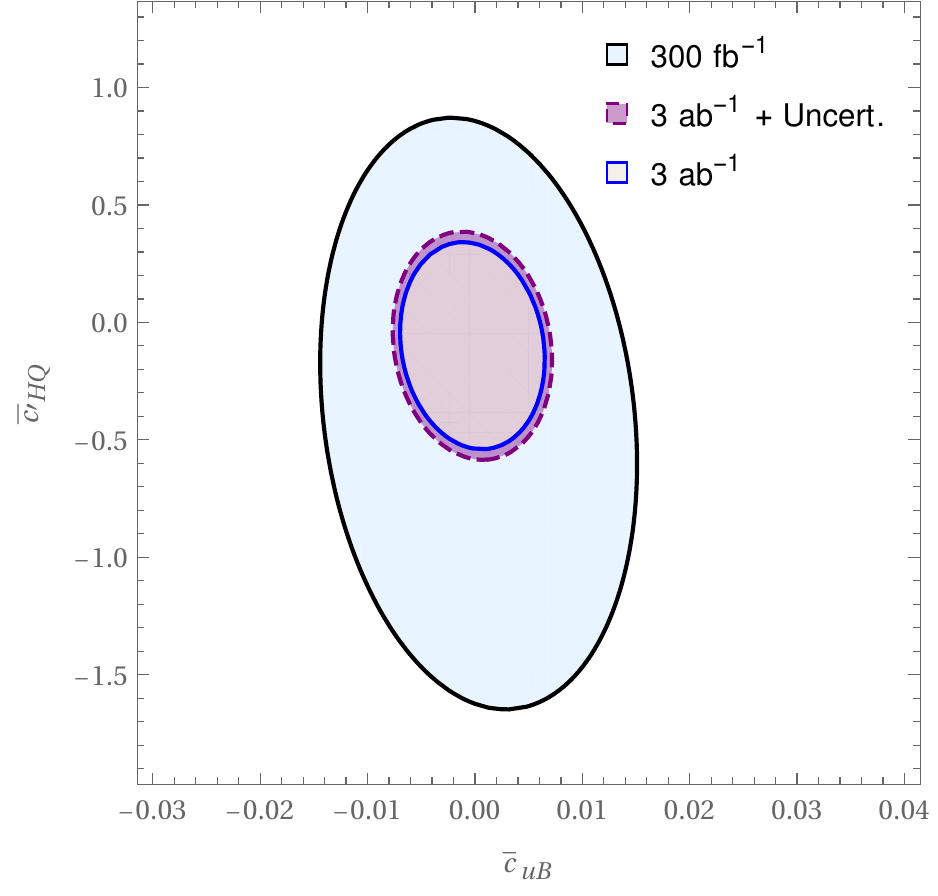}}
	\resizebox{0.23\textwidth}{!}{\includegraphics{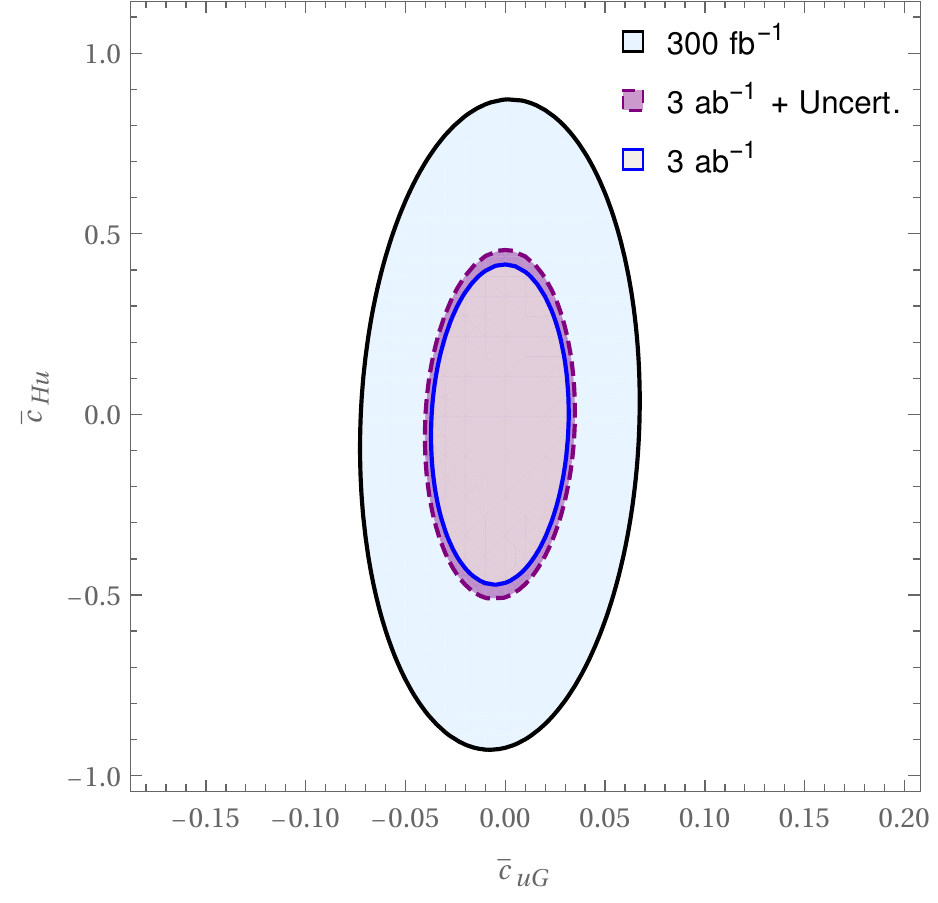}}
	\resizebox{0.23\textwidth}{!}{\includegraphics{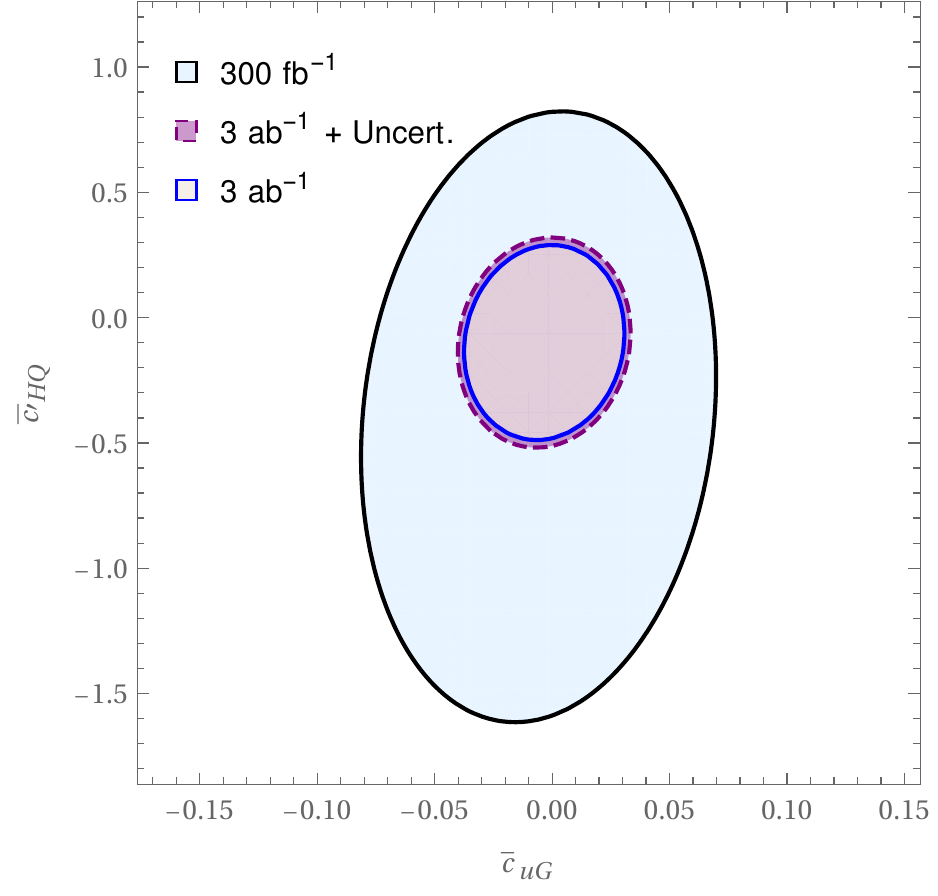}}
		\caption{{\small  Contours of 95\% CL at center-of-mass energy of 3000 GeV with the integrated luminosities of 300 and 3000 fb$^{-1}$.  The contours of 95\% CL
		considering an uncertainty of $10\%$ on the background rates and $10\%$ uncertainty on the signal efficiency with 3000 fb$^{-1}$.
		} \label{fig:Contours3000}}
	\end{center}
\end{figure*}
%------------------------------------------------

%------------------------------------------------
\begin{figure*}[htb]
	\vspace{0.50cm}	
	\begin{center}
		\resizebox{0.23\textwidth}{!}{\includegraphics{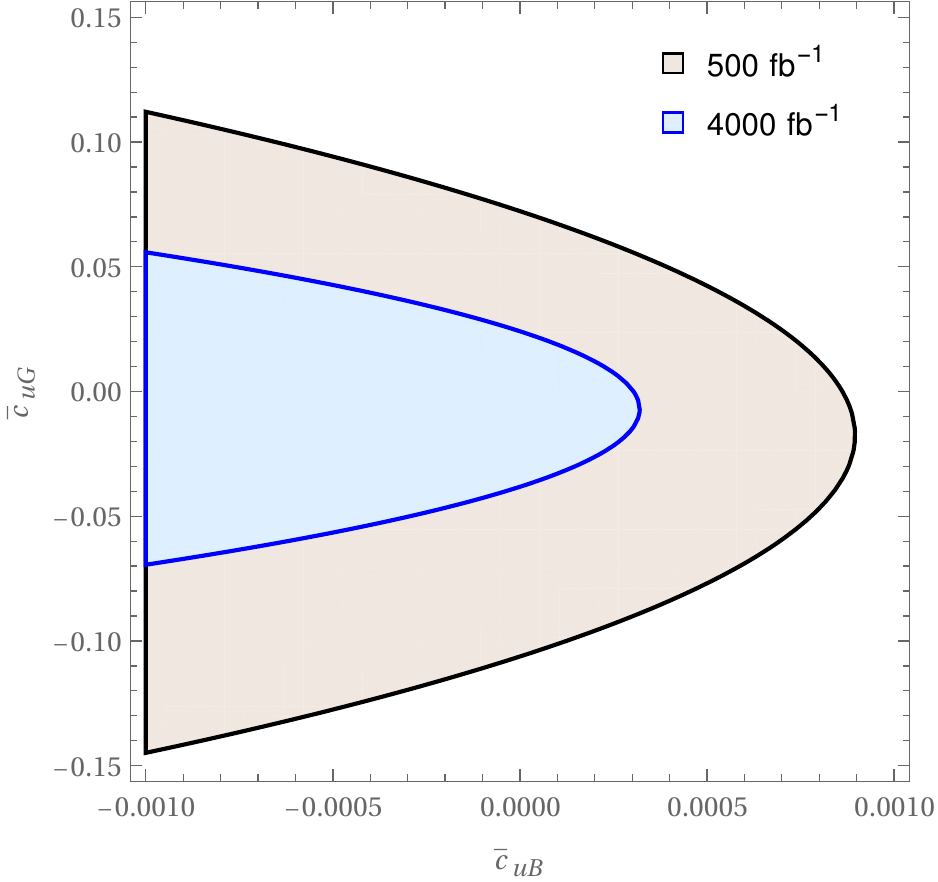}}
		\resizebox{0.23\textwidth}{!}{\includegraphics{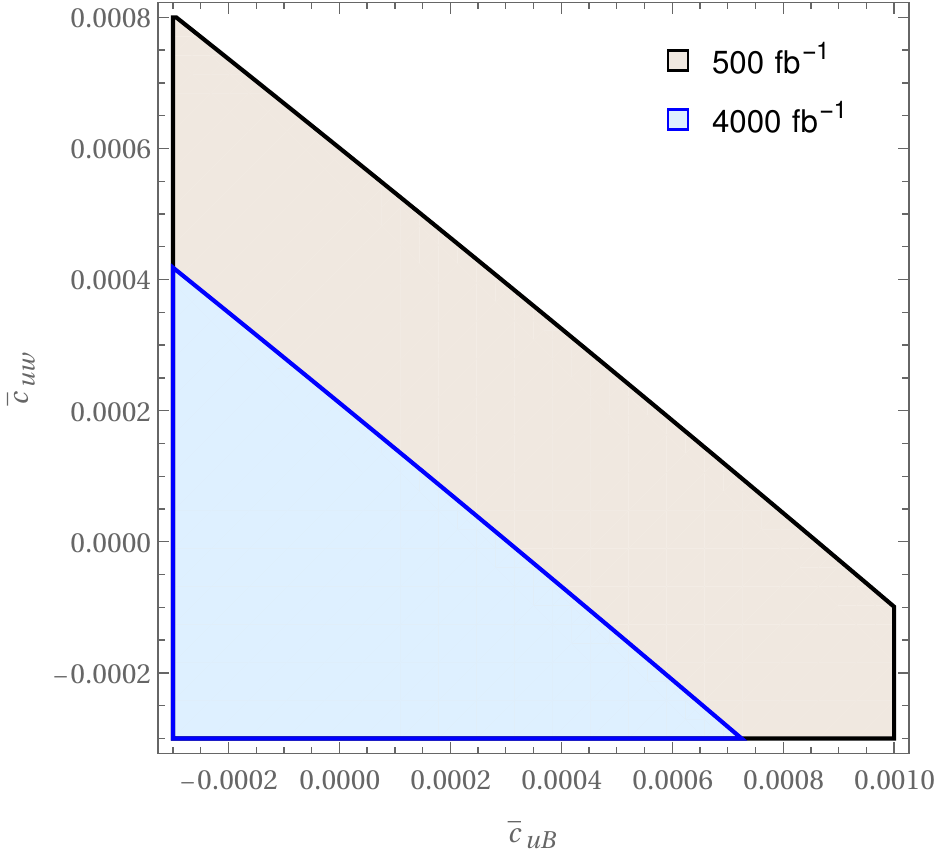}}
		\resizebox{0.23\textwidth}{!}{\includegraphics{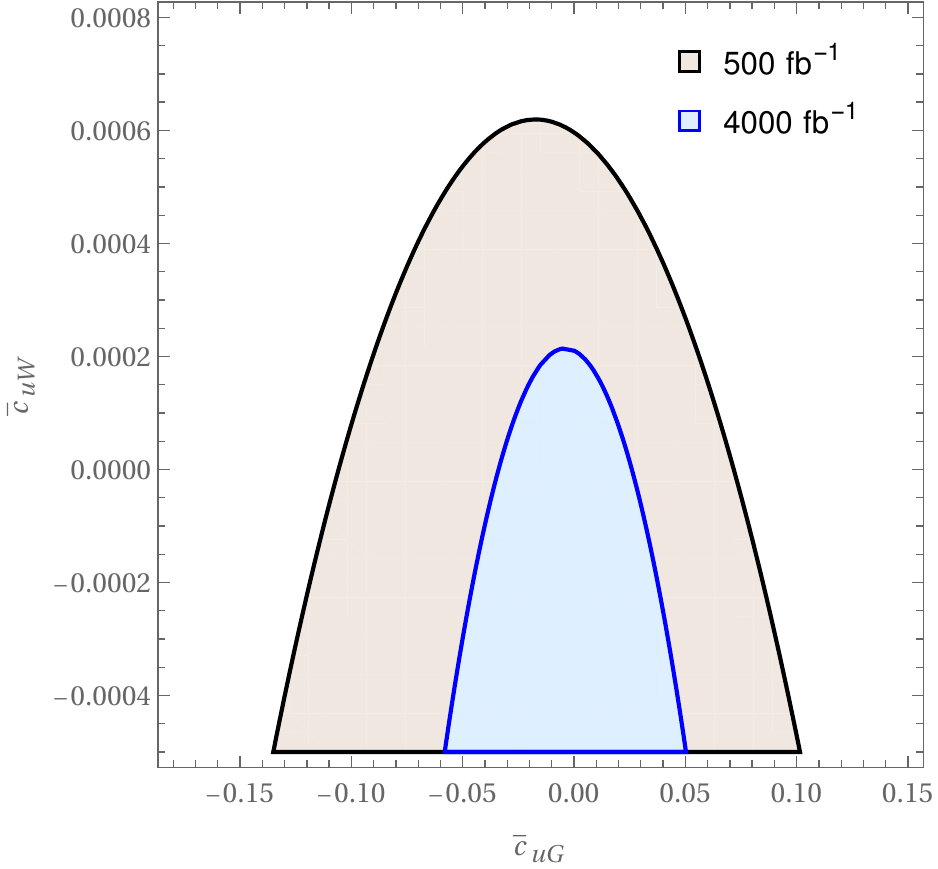}}
		\resizebox{0.22\textwidth}{!}{\includegraphics{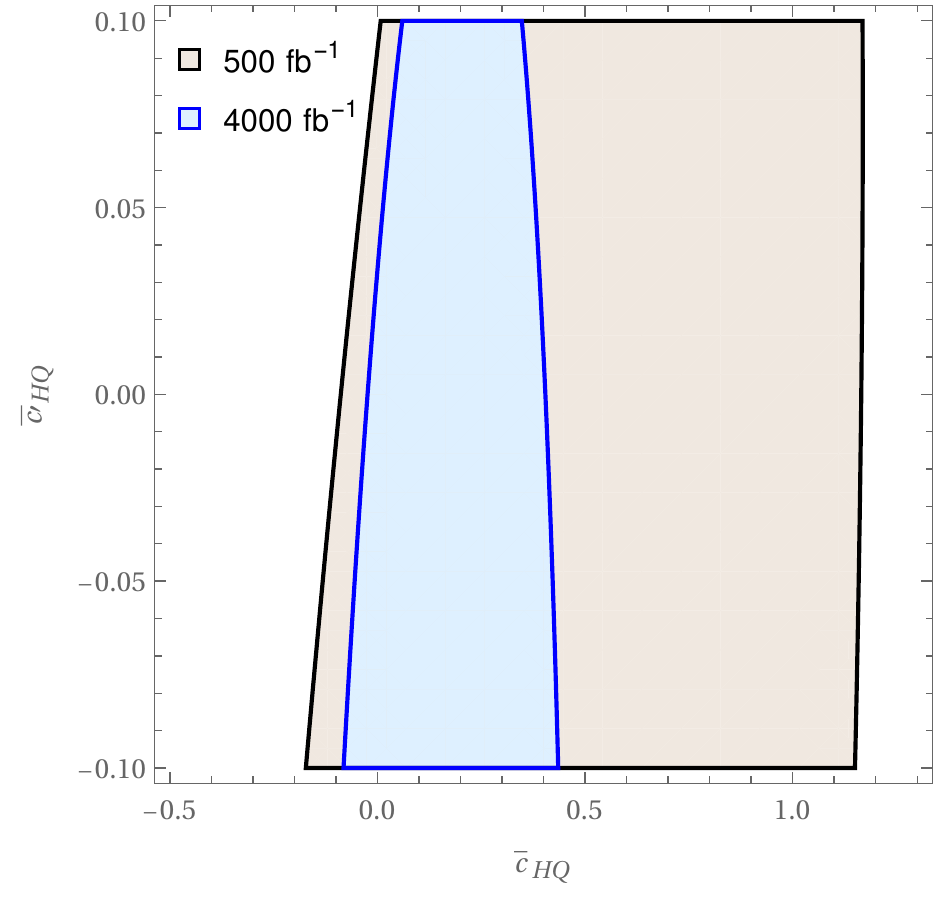}}
		\resizebox{0.23\textwidth}{!}{\includegraphics{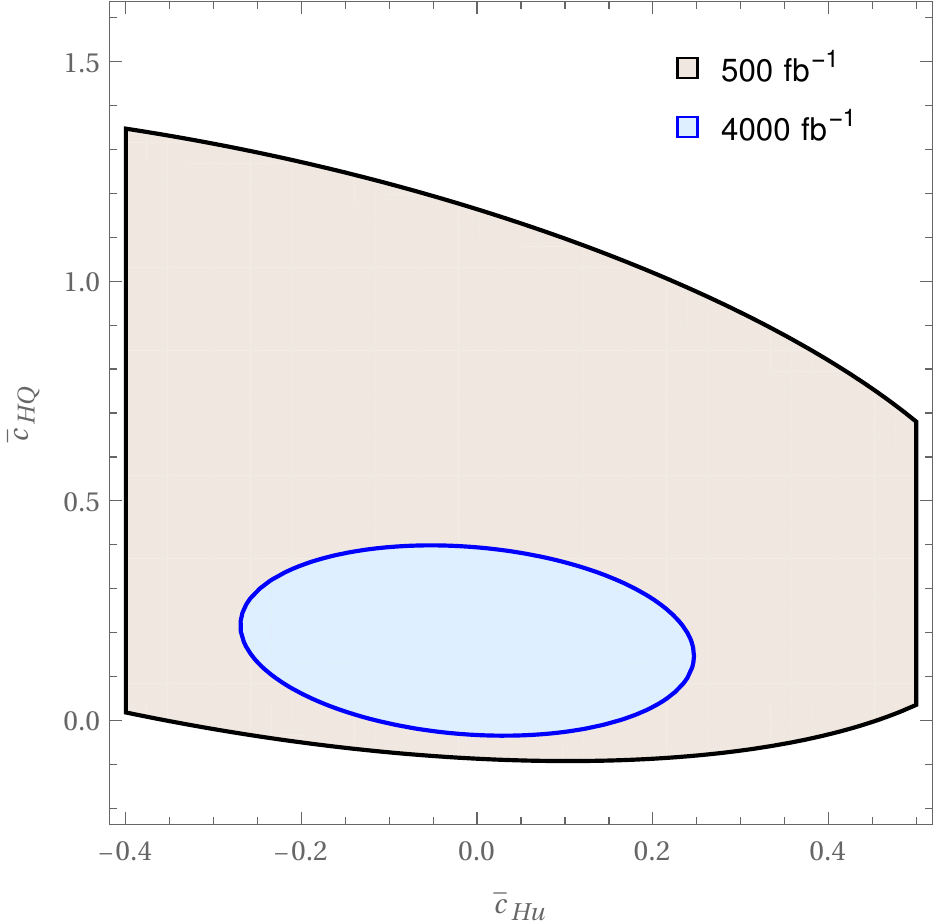}}
	\resizebox{0.23\textwidth}{!}{\includegraphics{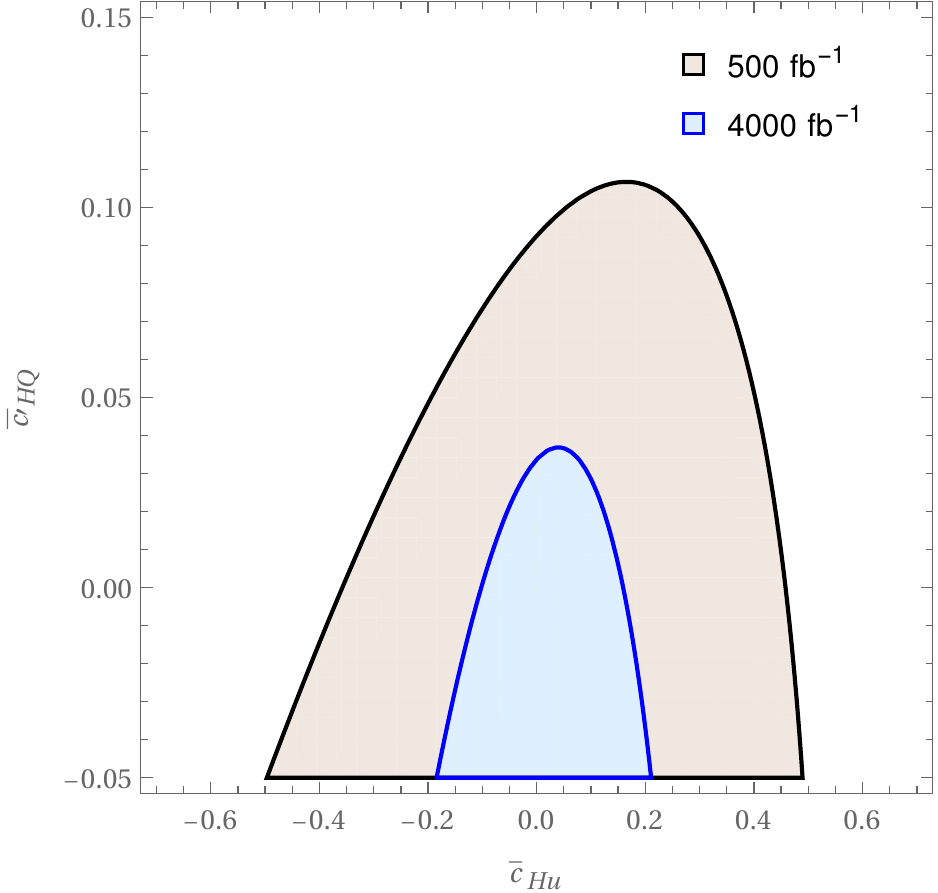}}
	\resizebox{0.23\textwidth}{!}{\includegraphics{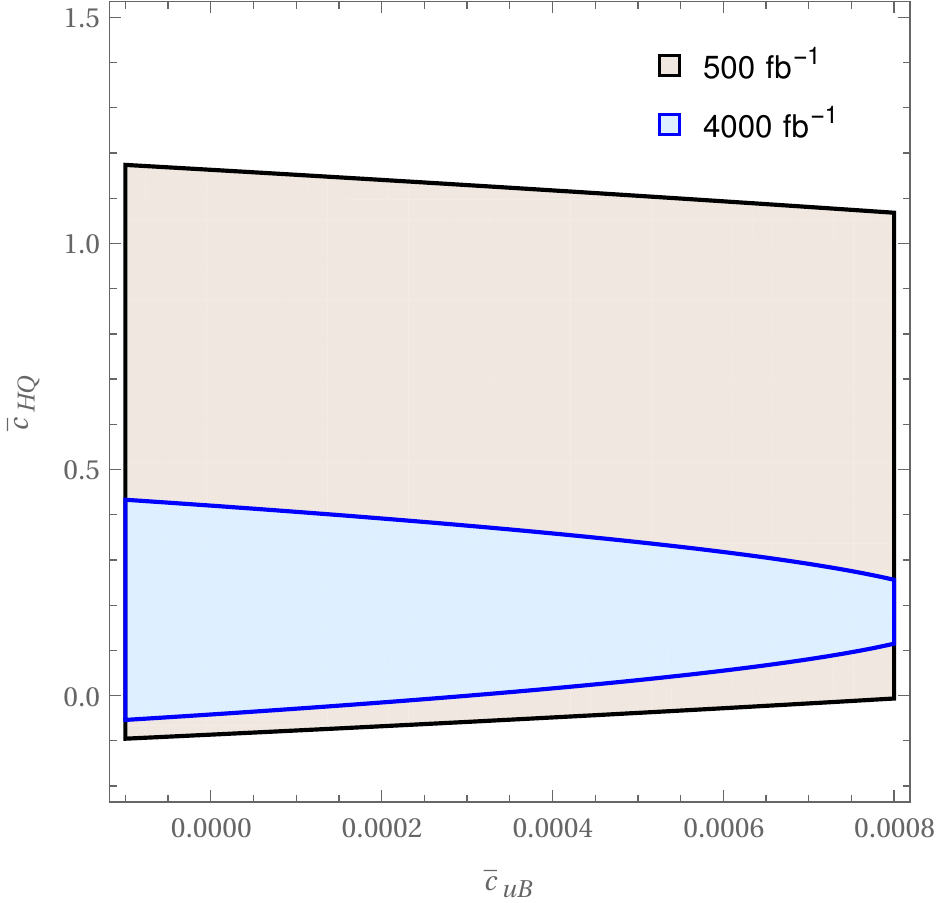}}
	\resizebox{0.23\textwidth}{!}{\includegraphics{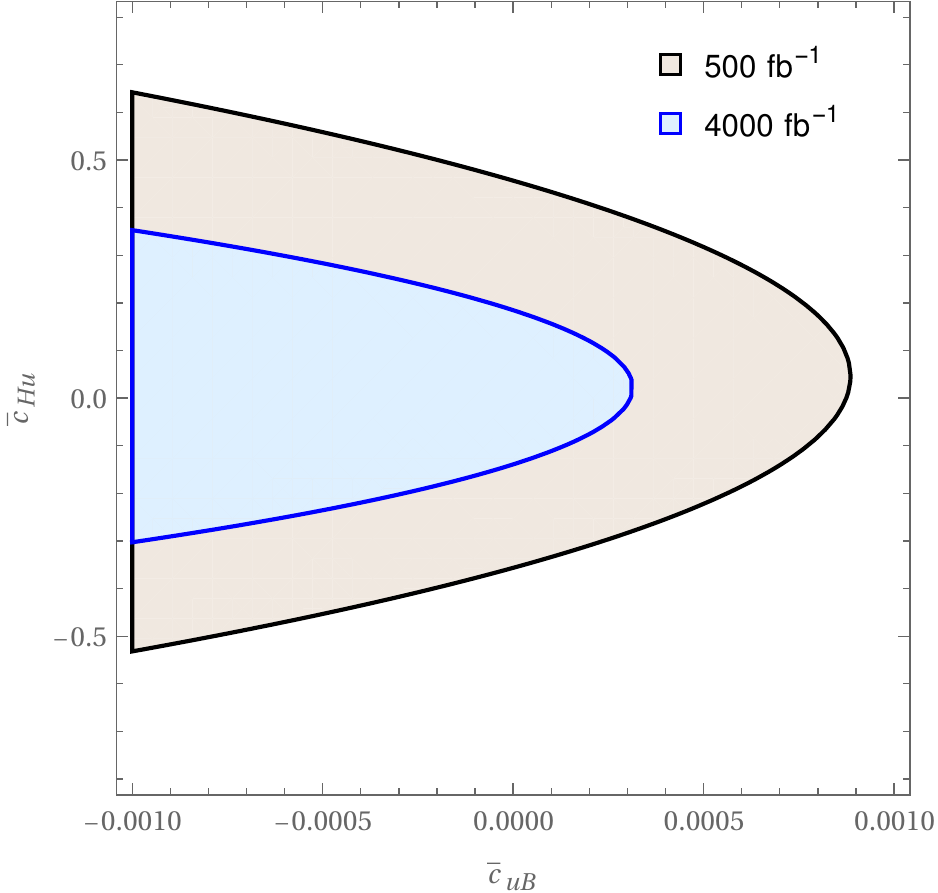}}
	\resizebox{0.23\textwidth}{!}{\includegraphics{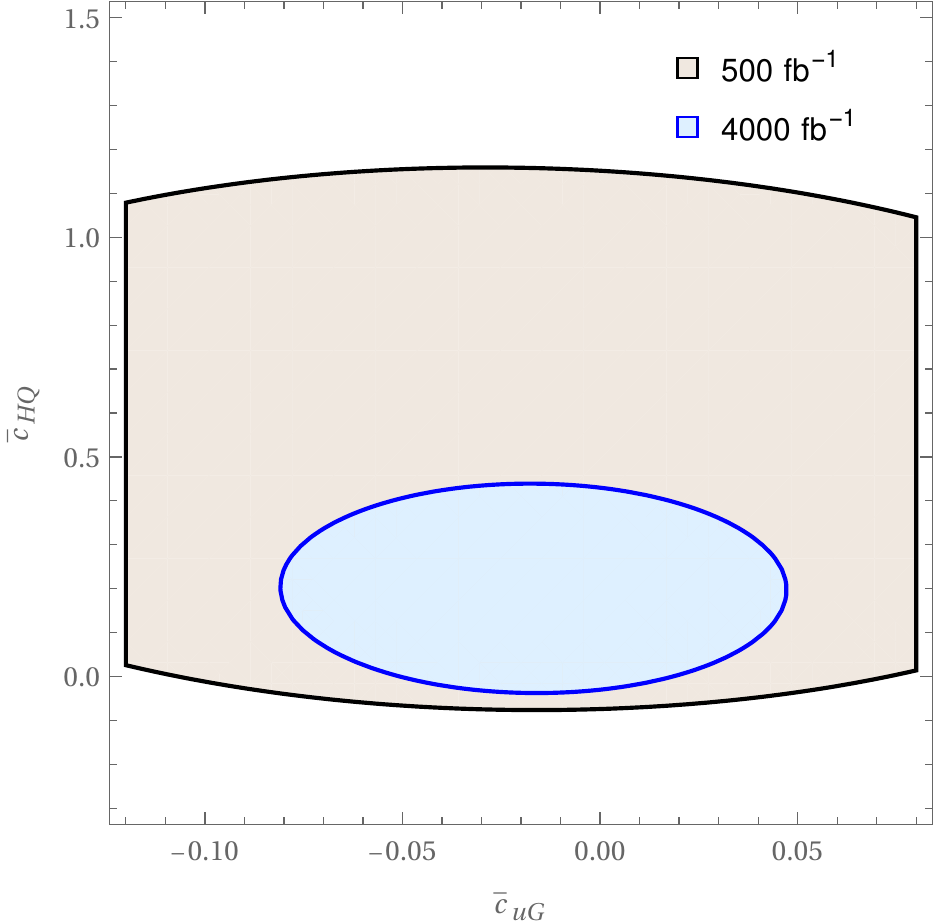}}
	\resizebox{0.23\textwidth}{!}{\includegraphics{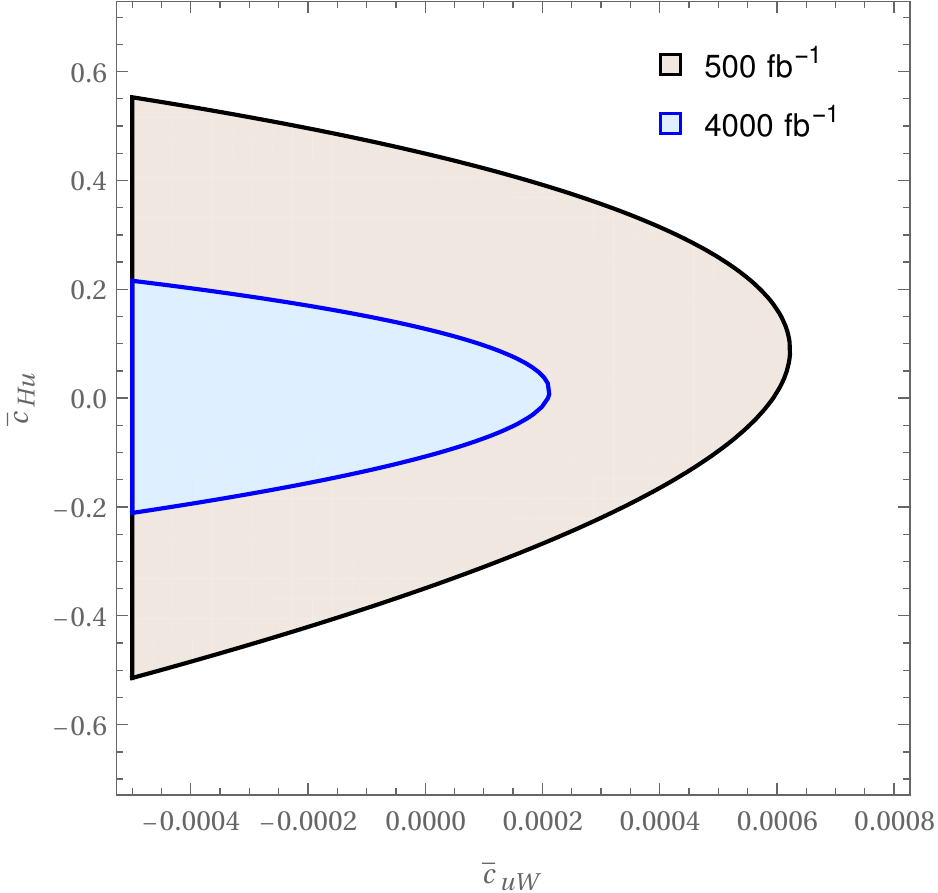}}
	\resizebox{0.23\textwidth}{!}{\includegraphics{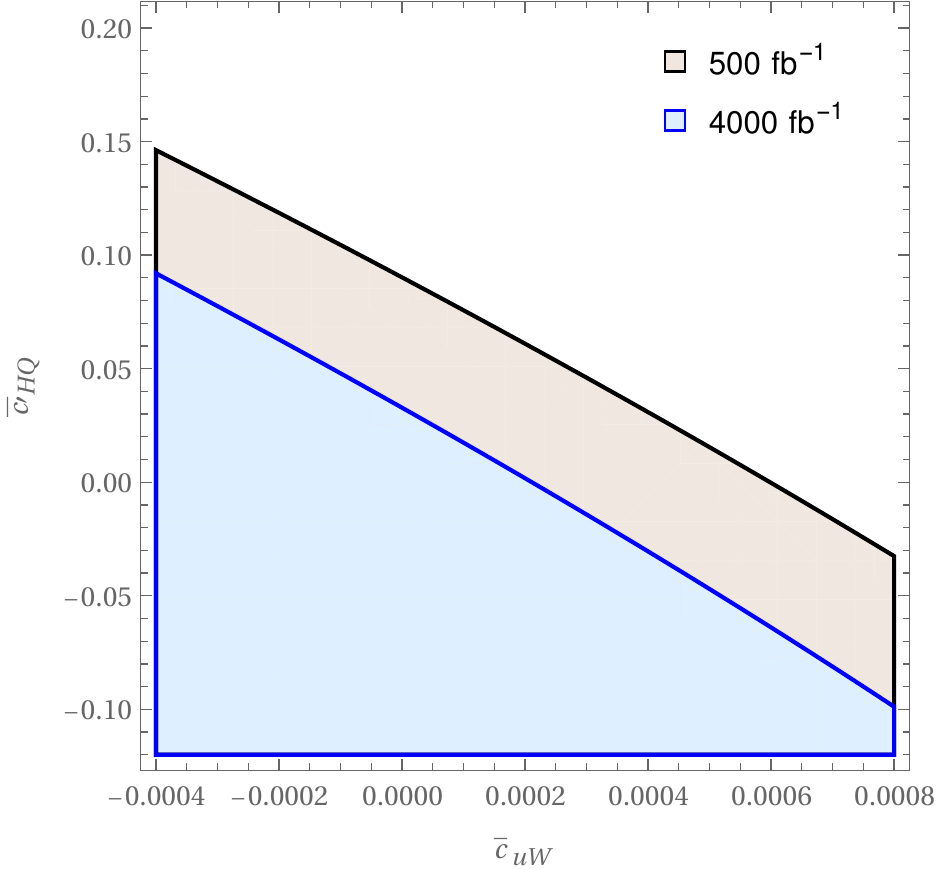}}
	\resizebox{0.23\textwidth}{!}{\includegraphics{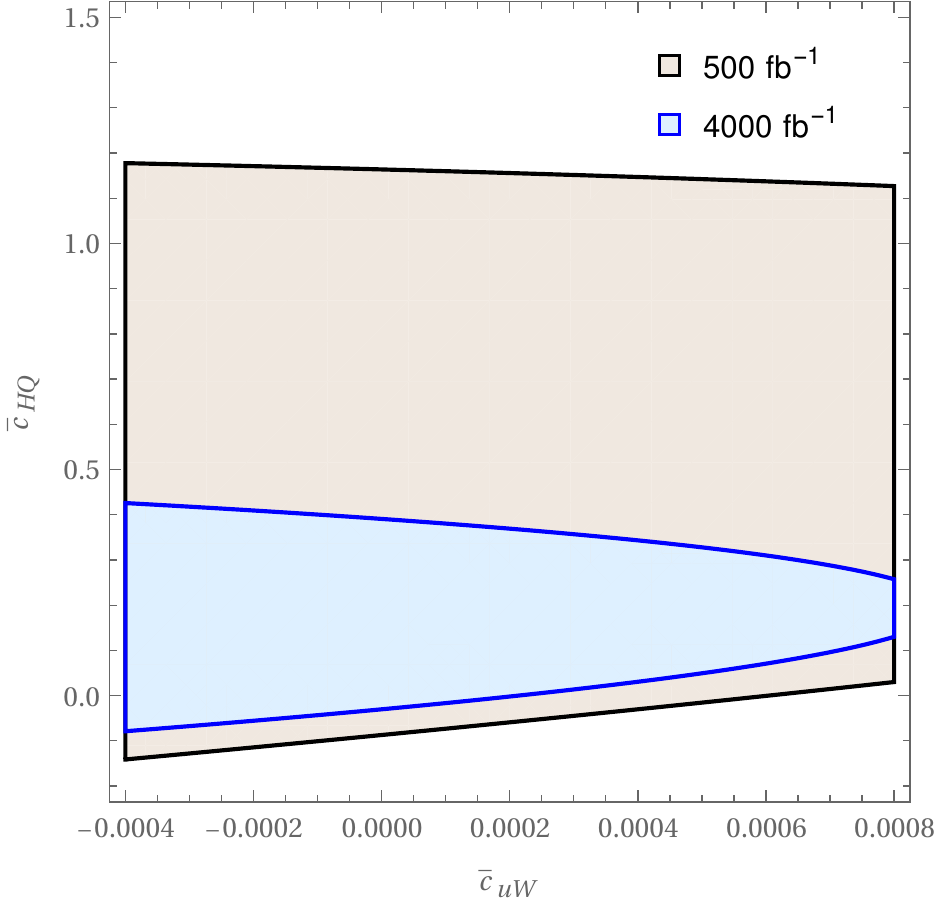}}
		\resizebox{0.23\textwidth}{!}{\includegraphics{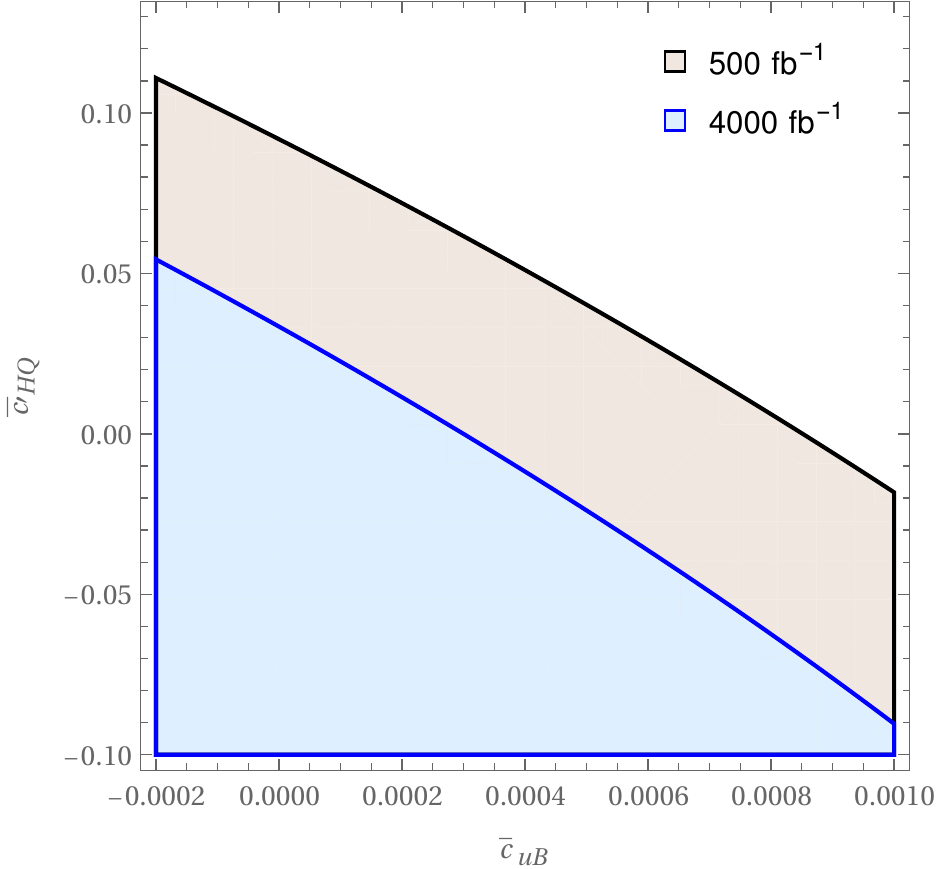}}
	\resizebox{0.22\textwidth}{!}{\includegraphics{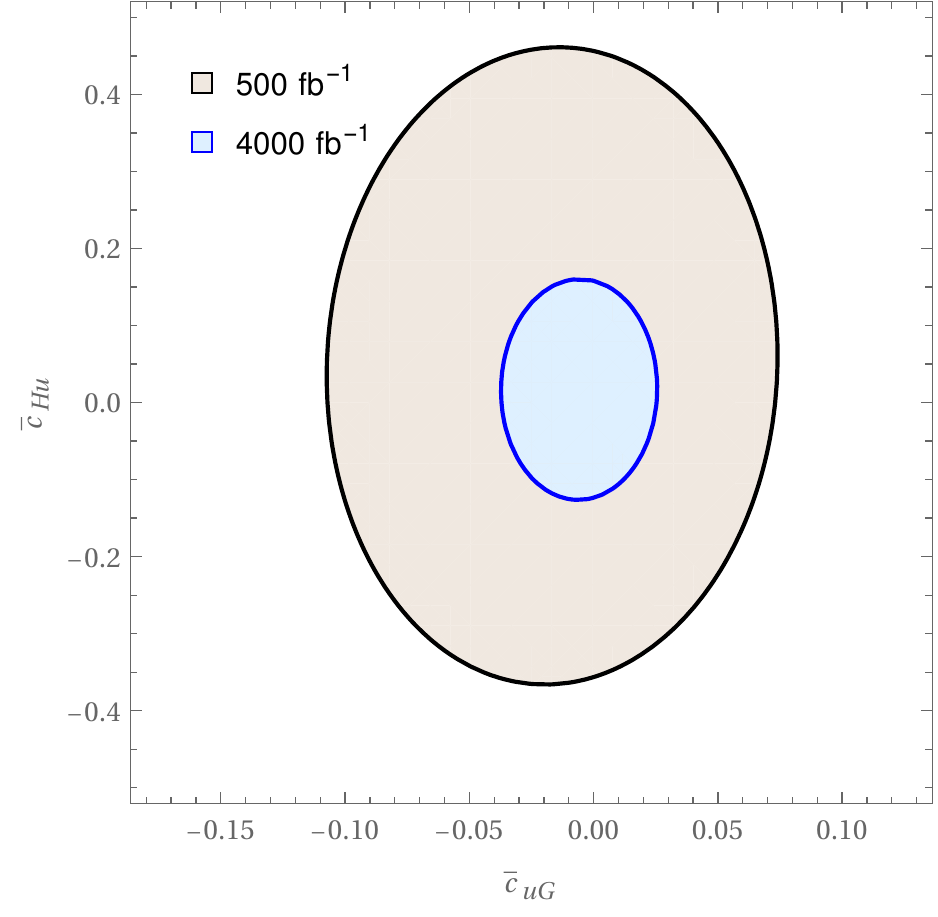}}
	\resizebox{0.23\textwidth}{!}{\includegraphics{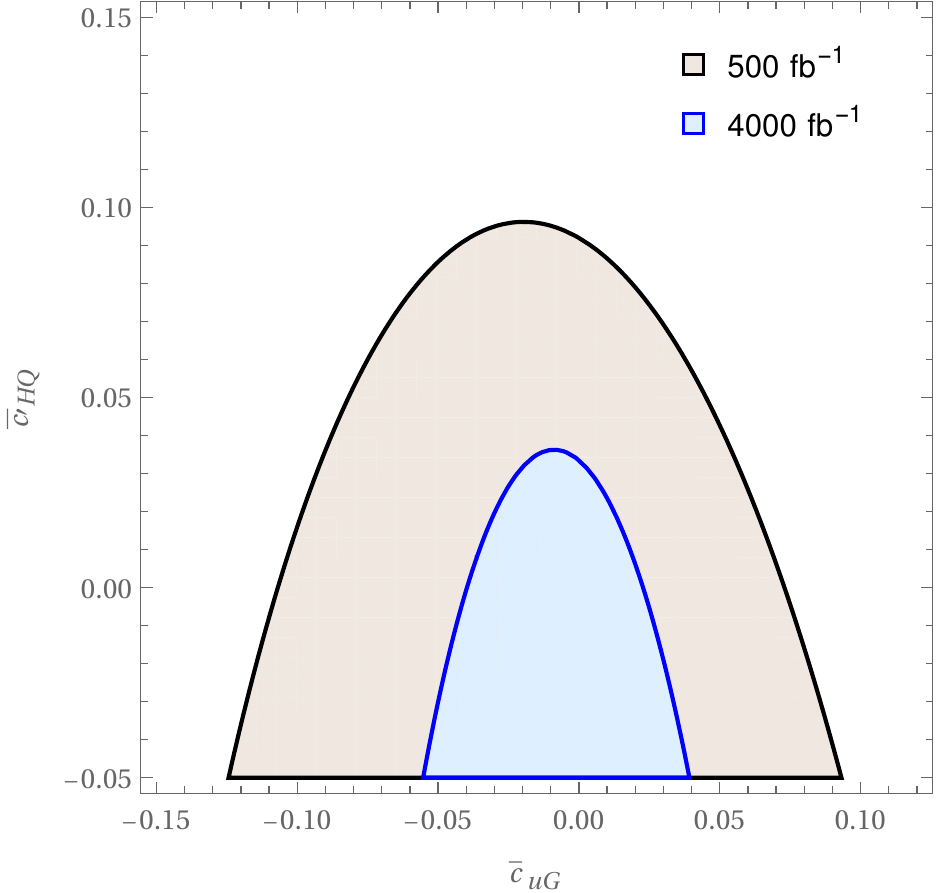}}
		\caption{{\small  Contours of 95\% confidence level at center-of-mass energy of 500 GeV with the integrated luminosities of 500  fb$^{-1}$ and 4000 fb$^{-1}$. 
		 Some plots are magnified for better clarity.} \label{fig:Contours500}}
	\end{center}
\end{figure*}
%------------------------------------------------

%--------------------------------
\begin{table}[ht]
	\begin{center} \small
		\caption{ The $95\%$ CL constraints on  the Wilson coefficients for different assumptions
			on the integrated luminosity at the center-of-mass energies of 500 and 3000 GeV.}
		\begin{tabular}{c|c|c|c|c}\hline
			Wilson coefficient & $ 500$ GeV, 500 fb$^{-1}$ & $ 500$ GeV, 4 ab$^{-1}$ & $ 3$ TeV, 300 fb$^{-1}$  & $ 3$ TeV, 3 ab$^{-1}$           \\ \hline 
			$\bar{c}_{uB} $ & $[-0.0476, 0.0009]$ & $[-0.017, 0.0003]$ & $[-0.013, 0.012]$	  & $[-0.0067, 0.0058]$    \\ \hline 
			$\bar{c}_{uG} $  & $[-0.11, 0.073]$ &	$[-0.039, 0.025]$ & $[-0.073, 0.068]$	  & $[-0.038, 0.033]$    \\ \hline 
			$\bar{c}_{uW} $  & $[-0.0294, 0.0006]$ &	 $[-0.011, 0.0002]$  & $[-0.0098, 0.0082]$ & $[-0.0051, 0.0035]$    \\ \hline    
			$\bar{c}_{Hu} $ & $[-0.35, 0.45]$ & $[-0.12, 0.16]$ & $[-1.00, 0.95]$	  & $[-0.51, 0.46]$    \\ \hline 
			$\bar{c}_{HQ} $  & $[-0.087, 1.17]$ &	$[-0.032, 0.41]$ & $[-1.21, 2.53]$	  & $[-0.37, 1.69]$    \\ \hline 
			$\bar{c}'_{HQ} $  & $[-1.34, 0.093]$ &	 $[-0.48, 0.034]$  & $[-1.63, 0.86]$ & $[-0.54, 0.31]$    \\ \hline  			     
		\end{tabular}
		\label{res2}
	\end{center}
\end{table}
%--------------------------------

From Table \ref{res2},  we see that the limits derived from the analysis of 3 ab$^{-1}$ of data from electron-positron collisions at 
$\sqrt{s} = 3$ TeV on the $\bar{c}_{uW}$ and $\bar{c}_{uB}$ are at the level of 
 $10^{-3}$, respectively.  
 For $\sqrt{s} = 500$ GeV, $\bar{c}_{uW}$ and $\bar{c}_{uB}$ reach one order better
sensitivity with the integrated luminosities of 500 fb$^{-1}$ and 4000 fb$^{-1}$.

 With the assumption of $c_{X} = 1$, 
the bound on $\bar{c}_{uW}$ (from $\sqrt{s} = 500$ GeV with 4000 fb$^{-1}$) corresponds to a mass scale of $\Lambda \gtrsim 17$ TeV.
The validity of the effective theory is determined by the energy scale of the process which is 
 fixed at lepton colliders and is equal to the center-of-mass energy.
The obtained constraints from this analysis are larger than the energy scale of the interaction,
 {\it i.e.} $\Lambda > \sqrt{s}$, which is consistent with the EFT description. 
The $\mathcal{O}_{uW}$ and $\mathcal{O}_{uB}$ operators have been probed 
using the $t\bar{t}$ production in Ref.\cite{Englert:2017dev}
for various scenarios at the CLIC and ILC.  
It has been shown that using observables such as total cross section,  forward-backward asymmetries, and 
utilising different sets of beam polarisation would lead to  constraints on $\bar{c}_{uW}$ and $\bar{c}_{uB}$ at the order of $\lesssim 10^{-4}$.
The results from this analysis
derives  comparable bounds on $\mathcal{O}_{uW}$ and $\mathcal{O}_{uB}$ operators
with those from Ref.\cite{Englert:2017dev}. We note that 
combining the semi-leptonic topology of $t\bar{t}j$ process with the dileptonic one, considered in this analysis, would improve the bounds. 

The derived limits in this analysis could be used to probe the parameters of  explicit models  which their low energy limits
tend to the SMEFT. For instance, in beyond the SM scenarios with strongly interacting
Higgs boson, a naive estimation leads to the following  for the Wilson coefficients \cite{Contino:2013kra,Giudice:2007fh}:
\begin{eqnarray}
\bar{c}_{uW}, ~\bar{c}_{uB},~ \bar{c}_{uG} \sim \mathcal{O}\big(\frac{g^{*2} m_{W}^{2}}{16\pi^2M^{2}} \big).
\end{eqnarray}
where $M$ is the mass scale of the new physical state and $g^{*} (\leq 4\pi)$ denotes the coupling strength of the Higgs boson to the new
physics state. The obtained limit on $\bar{c}_{uW}$ at $\sqrt{s} = 3$ TeV with 3 ab$^{-1}$ integrated luminosity of data lead to 
a lower bound of $7$ TeV on $M$, in the strongly interacting regime $g^{*} = 4 \pi$.

}
%
%%%%%%%%%%%%%%%%%%%%%%%%%%%%%%%%%%%%%%%%%%%%%%%%%%%%%%%%%%%%%%%%%%%%%%
\section{Summary and conclusions} \label{sec:Discussion}
%%%%%%%%%%%%%%%%%%%%%%%%%%%%%%%%%%%%%%%%%%%%%%%%%%%%%%%%%%%%%%%%%%%%%%
{

We  perform a study to probe the sensitivity of future lepton colliders to the top quark effective 
 couplings at the center-of-mass energies of 500 and 3000 GeV.
In particular, we concentrate on the top pair production in association with a jet
within the SMEFT framework. The SMEFT is an attractive and an efficient way to describe the possible effects of new physics 
until new particles from beyond the SM are observed. The $e^{-} e^{+} \rightarrow t \bar{t}$+jet process
is found to be mostly sensitive to $\mathcal{O}_{uW}$ and $\mathcal{O}_{uB}$ operators, respectively. 
The clean environment at lepton colliders and the expected high resolution for measurements of leptons and
jets properties allow us  to characterise the $t\bar{t}+$jet events through 
the dileptonic channel, where the final state consists of two charged lepton ($\ell^\pm$), at least two jets from which 
two are originating from hadronisation of $b$-quarks, and missing transverse momentum.
The results are based on a comprehensive analysis where  the major sources of background processes and a realistic simulation 
of the detector response, flavour tagging, and jet clustering have been considered.
A set of kinematic variables consisting of scalar sum of the transverse momentum of the
leptons and jets, invariant mass of the b-jets, and pseudorapidity of the leading lepton and b-jet are used
as  input  to a multivariate analysis for separation of signal from background processes.

It is found that using lepton colliders at both center-of-mass energies  $\sqrt{s} = 3000$ and 500 GeV
 would allow us to constrain the $\bar{c}_{uW}$ and $\bar{c}_{uB}$ of the order of 
$10^{-3}$ and $10^{-4}$, respectively.  

}

%
%%%%%%%%%%%%%%%%%%%%%%%%%%%%%%%%%%%%%%%%%%%%%%%%%%%%%%%%%%%%%%%%%%%%%%
\section*{Acknowledgments}
%%%%%%%%%%%%%%%%%%%%%%%%%%%%%%%%%%%%%%%%%%%%%%%%%%%%%%%%%%%%%%%%%%%%%%
{	
We are grateful to MadAnalysis and MatchChecker authors and Pedro Vieira De Castro Ferreira Da Silva for answering our questions
related to merging. 
We also thank F. Elahi and S. M. Etesami for useful comments and fruitful
discussions. Hamzeh Khanpour is thankful to the University of Science and Technology 
of Mazandaran for financial support provided for this project and 
is grateful to the CERN theory department for their hospitality and support during the preparation of this paper. 
}

\clearpage

%
%
%%%%%%%%%%%%%%%%%%%%%%%%%%%%%%%%


\begin{thebibliography}{}
%%%%%%%%%%%%%%%%%%%%%%%%%%%%%%%%
%

%\cite{Aad:2012tfa}
\bibitem{Aad:2012tfa} 
G.~Aad {\it et al.} [ATLAS Collaboration],
  ``Observation of a new particle in the search for the Standard Model Higgs boson with the ATLAS detector at the LHC,''
  Phys.\ Lett.\ B {\bf 716}, 1 (2012)
  doi:10.1016/j.physletb.2012.08.020
  [arXiv:1207.7214 [hep-ex]].
  %%CITATION = doi:10.1016/j.physletb.2012.08.020;%%
  %9644 citations counted in INSPIRE as of 27 Aug 2019


%\cite{Chatrchyan:2012xdj}
\bibitem{Chatrchyan:2012xdj} 
 S.~Chatrchyan {\it et al.} [CMS Collaboration],
  ``Observation of a New Boson at a Mass of 125 GeV with the CMS Experiment at the LHC,''
  Phys.\ Lett.\ B {\bf 716}, 30 (2012)
  doi:10.1016/j.physletb.2012.08.021
  [arXiv:1207.7235 [hep-ex]].
  %%CITATION = doi:10.1016/j.physletb.2012.08.021;%%
  %9432 citations counted in INSPIRE as of 27 Aug 2019




%\cite{Dawson:2013bba}
\bibitem{Dawson:2013bba} 
 S.~Dawson {\it et al.},
  ``Working Group Report: Higgs Boson,''
  arXiv:1310.8361 [hep-ex].
  %%CITATION = ARXIV:1310.8361;%%
  %389 citations counted in INSPIRE as of 27 Aug 2019



%\cite{deFlorian:2016spz}
\bibitem{deFlorian:2016spz} 
 D.~de Florian {\it et al.} [LHC Higgs Cross Section Working Group],
  ``Handbook of LHC Higgs Cross Sections: 4. Deciphering the Nature of the Higgs Sector,''
  doi:10.23731/CYRM-2017-002
  arXiv:1610.07922 [hep-ph].
  %%CITATION = doi:10.23731/CYRM-2017-002;%%
  %736 citations counted in INSPIRE as of 27 Aug 2019



%\cite{Heinemeyer:2013tqa}
\bibitem{Heinemeyer:2013tqa} 
 S.~Heinemeyer {\it et al.} [LHC Higgs Cross Section Working Group],
  ``Handbook of LHC Higgs Cross Sections: 3. Higgs Properties,''
  doi:10.5170/CERN-2013-004
  arXiv:1307.1347 [hep-ph].
  %%CITATION = doi:10.5170/CERN-2013-004;%%
  %1231 citations counted in INSPIRE as of 27 Aug 2019




%\cite{Abe:1995hr}
\bibitem{Abe:1995hr} 
  F.~Abe {\it et al.} [CDF Collaboration],
  ``Observation of top quark production in $\bar{p}p$ collisions,''
  Phys.\ Rev.\ Lett.\  {\bf 74}, 2626 (1995)
  doi:10.1103/PhysRevLett.74.2626
  [hep-ex/9503002].
  %%CITATION = doi:10.1103/PhysRevLett.74.2626;%%
  %3324 citations counted in INSPIRE as of 27 Aug 2019
  
  
%\cite{Abachi:1994td}
\bibitem{Abachi:1994td} 
 S.~Abachi {\it et al.} [D0 Collaboration],
  ``Search for high mass top quark production in $p\bar{p}$ collisions at $\sqrt{s} = 1.8$ TeV,''
  Phys.\ Rev.\ Lett.\  {\bf 74}, 2422 (1995)
  doi:10.1103/PhysRevLett.74.2422
  [hep-ex/9411001].
  %%CITATION = doi:10.1103/PhysRevLett.74.2422;%%
  %341 citations counted in INSPIRE as of 27 Aug 2019


%\cite{Fan:2014vta}
\bibitem{Fan:2014vta}
 J.~Fan, M.~Reece and L.~T.~Wang,
  ``Possible Futures of Electroweak Precision: ILC, FCC-ee, and CEPC,''
  JHEP {\bf 1509}, 196 (2015)
  doi:10.1007/JHEP09(2015)196
  [arXiv:1411.1054 [hep-ph]].
  %%CITATION = doi:10.1007/JHEP09(2015)196;%%
  %59 citations counted in INSPIRE as of 27 Aug 2019

%\cite{Moortgat-Picka:2015yla}
\bibitem{Moortgat-Picka:2015yla}
 G.~Moortgat-Pick {\it et al.},
  ``Physics at the e+ e- Linear Collider,''
  Eur.\ Phys.\ J.\ C {\bf 75}, no. 8, 371 (2015)
  doi:10.1140/epjc/s10052-015-3511-9
  [arXiv:1504.01726 [hep-ph]].
  %%CITATION = doi:10.1140/epjc/s10052-015-3511-9;%%
  %141 citations counted in INSPIRE as of 27 Aug 2019


%\cite{Fujii:2015jha}
\bibitem{Fujii:2015jha}
 K.~Fujii {\it et al.},
  ``Physics Case for the International Linear Collider,''
  arXiv:1506.05992 [hep-ex].
  %%CITATION = ARXIV:1506.05992;%%
  %183 citations counted in INSPIRE as of 27 Aug 2019

%\cite{Charles:2018vfv}
\bibitem{Charles:2018vfv}
 P.~N.~Burrows {\it et al.} [CLICdp and CLIC Collaborations],
  ``The Compact Linear Collider (CLIC) - 2018 Summary Report,''
  CERN Yellow Rep.\ Monogr.\  {\bf 1802}, 1 (2018)
  doi:10.23731/CYRM-2018-002
  [arXiv:1812.06018 [physics.acc-ph]].
  %%CITATION = doi:10.23731/CYRM-2018-002;%%
  %24 citations counted in INSPIRE as of 27 Aug 2019


%\cite{Roloff:2018dqu}
\bibitem{Roloff:2018dqu}
P.~Roloff {\it et al.} [CLIC and CLICdp Collaborations],
  ``The Compact Linear e$^+$e$^-$ Collider (CLIC): Physics Potential,''
  arXiv:1812.07986 [hep-ex].
  %%CITATION = ARXIV:1812.07986;%%
  %6 citations counted in INSPIRE as of 27 Aug 2019
  
  
%\cite{Aicheler:2019dhf}
\bibitem{Aicheler:2019dhf}
 M.~Aicheler {\it et al.} [CLIC accelerator Collaboration],
  ``The Compact Linear Collider (CLIC) - Project Implementation Plan,''
  doi:10.23731/CYRM-2018-004
  arXiv:1903.08655 [physics.acc-ph].
  %%CITATION = doi:10.23731/CYRM-2018-004;%%
  %8 citations counted in INSPIRE as of 27 Aug 2019
  
  
%\cite{Aihara:2019gcq}
\bibitem{Aihara:2019gcq}
 H.~Aihara {\it et al.} [ILC Collaboration],
  ``The International Linear Collider. A Global Project,''
  arXiv:1901.09829 [hep-ex].
  %%CITATION = ARXIV:1901.09829;%%
  %9 citations counted in INSPIRE as of 27 Aug 2019



%\cite{Baer:2013cma}
\bibitem{Baer:2013cma}
 H.~Baer {\it et al.},
  ``The International Linear Collider Technical Design Report - Volume 2: Physics,''
  arXiv:1306.6352 [hep-ph].
  %%CITATION = ARXIV:1306.6352;%%
  %750 citations counted in INSPIRE as of 27 Aug 2019




%\cite{Behnke:2013xla}
\bibitem{Behnke:2013xla}
T.~Behnke {\it et al.},
  ``The International Linear Collider Technical Design Report - Volume 1: Executive Summary,''
  arXiv:1306.6327 [physics.acc-ph].
  %%CITATION = ARXIV:1306.6327;%%
  %338 citations counted in INSPIRE as of 27 Aug 2019




%\cite{CEPC-SPPCStudyGroup:2015csa}
\bibitem{CEPC-SPPCStudyGroup:2015csa}
  M.~Ahmad {\it et al.},
  ``CEPC-SPPC Preliminary Conceptual Design Report. 1. Physics and Detector,''
  IHEP-CEPC-DR-2015-01, IHEP-TH-2015-01, IHEP-EP-2015-01.
  %%CITATION = IHEP-CEPC-DR-2015-01, IHEP-TH-2015-01, IHEP-EP-2015-01;%%
  %186 citations counted in INSPIRE as of 27 Aug 2019


%\cite{CEPC-SPPCStudyGroup:2015esa}
\bibitem{CEPC-SPPCStudyGroup:2015esa}
  CEPC-SPPC Study Group,
  ``CEPC-SPPC Preliminary Conceptual Design Report. 2.  Accelerator,''
  IHEP-CEPC-DR-2015-01, IHEP-AC-2015-01.
  %%CITATION = IHEP-CEPC-DR-2015-01, IHEP-AC-2015-01;%%
  %100 citations counted in INSPIRE as of 27 Aug 2019


%\cite{Benedikt:2018qee}
\bibitem{Benedikt:2018qee}
 A.~Abada {\it et al.} [FCC Collaboration],
  ``FCC-ee: The Lepton Collider : Future Circular Collider Conceptual Design Report Volume 2,''
  Eur.\ Phys.\ J.\ ST {\bf 228}, no. 2, 261 (2019).
  doi:10.1140/epjst/e2019-900045-4
  %%CITATION = doi:10.1140/epjst/e2019-900045-4;%%
  %55 citations counted in INSPIRE as of 27 Aug 2019



%\cite{Gomez-Ceballos:2013zzn}
\bibitem{Gomez-Ceballos:2013zzn}
 M.~Bicer {\it et al.} [TLEP Design Study Working Group],
  ``First Look at the Physics Case of TLEP,''
  JHEP {\bf 1401}, 164 (2014)
  doi:10.1007/JHEP01(2014)164
  [arXiv:1308.6176 [hep-ex]].
  %%CITATION = doi:10.1007/JHEP01(2014)164;%%
  %538 citations counted in INSPIRE as of 27 Aug 2019





%\cite{Abada:2019lih}
\bibitem{Abada:2019lih} 
 A.~Abada {\it et al.} [FCC Collaboration],
  ``FCC Physics Opportunities : Future Circular Collider Conceptual Design Report Volume 1,''
  Eur.\ Phys.\ J.\ C {\bf 79}, no. 6, 474 (2019).
  doi:10.1140/epjc/s10052-019-6904-3
  %%CITATION = doi:10.1140/epjc/s10052-019-6904-3;%%
  %45 citations counted in INSPIRE as of 27 Aug 2019


%\cite{Abada:2019zxq}
\bibitem{Abada:2019zxq} 
 A.~Abada {\it et al.} [FCC Collaboration],
  ``FCC-ee: The Lepton Collider : Future Circular Collider Conceptual Design Report Volume 2,''
  Eur.\ Phys.\ J.\ ST {\bf 228}, no. 2, 261 (2019).
  doi:10.1140/epjst/e2019-900045-4
  %%CITATION = doi:10.1140/epjst/e2019-900045-4;%%
  %55 citations counted in INSPIRE as of 27 Aug 2019

%\cite{Azzi:2019yne}
\bibitem{Azzi:2019yne}
  P.~Azzi {\it et al.} [HL-LHC Collaboration and HE-LHC Working Group],
  ``Standard Model Physics at the HL-LHC and HE-LHC,''
  arXiv:1902.04070 [hep-ph].
  %%CITATION = ARXIV:1902.04070;%%
  %33 citations counted in INSPIRE as of 27 Aug 2019
  
  
%\cite{Beacham:2019nyx}
\bibitem{Beacham:2019nyx}
  J.~Beacham {\it et al.},
  ``Physics Beyond Colliders at CERN: Beyond the Standard Model Working Group Report,''
  arXiv:1901.09966 [hep-ex].
  %%CITATION = ARXIV:1901.09966;%%
  %31 citations counted in INSPIRE as of 27 Aug 2019

%\cite{CidVidal:2018eel}
\bibitem{CidVidal:2018eel}
 X.~Cid Vidal {\it et al.} [Working Group 3],
  ``Beyond the Standard Model Physics at the HL-LHC and HE-LHC,''
  arXiv:1812.07831 [hep-ph].
  %%CITATION = ARXIV:1812.07831;%%
  %35 citations counted in INSPIRE as of 27 Aug 2019


\bibitem{e2}
 F.~Maltoni, L.~Mantani and K.~Mimasu,
  ``Top-quark electroweak interactions at high energy,''
  arXiv:1904.05637 [hep-ph].
  %%CITATION = ARXIV:1904.05637;%%
  %2 citations counted in INSPIRE as of 06 Aug 2019


\bibitem{e3}
 C.~W.~Murphy,
  ``Statistical approach to Higgs boson couplings in the standard model effective field theory,''
  Phys.\ Rev.\ D {\bf 97}, no. 1, 015007 (2018)
  doi:10.1103/PhysRevD.97.015007
  [arXiv:1710.02008 [hep-ph]].
  %%CITATION = doi:10.1103/PhysRevD.97.015007;%%
  %13 citations counted in INSPIRE as of 06 Aug 2019

\bibitem{e4}
  G.~Durieux, M.~Perello, M.~Vos and C.~Zhang,
  ``Global and optimal probes for the top-quark effective field theory at future lepton colliders,''
  JHEP {\bf 1810}, 168 (2018)
  doi:10.1007/JHEP10(2018)168
  [arXiv:1807.02121 [hep-ph]].
  %%CITATION = doi:10.1007/JHEP10(2018)168;%%
  %15 citations counted in INSPIRE as of 06 Aug 2019


\bibitem{e5}
  G.~Durieux, J.~Gu, E.~Vryonidou and C.~Zhang,
  ``Probing top-quark couplings indirectly at Higgs factories,''
  Chin.\ Phys.\ C {\bf 42}, no. 12, 123107 (2018)
  doi:10.1088/1674-1137/42/12/123107
  [arXiv:1809.03520 [hep-ph]].
  %%CITATION = doi:10.1088/1674-1137/42/12/123107;%%
  %9 citations counted in INSPIRE as of 06 Aug 2019

\bibitem{e6}
  G.~Durieux, A.~Irles, V.~Miralles, A.~Penuelas, R.~Poschl, M.~Perello and M.~Vos,
  ``The electro-weak couplings of the top and bottom quarks -- global fit and future prospects,''
  arXiv:1907.10619 [hep-ph].
  %%CITATION = ARXIV:1907.10619;%%

\bibitem{e7}
M.~Chala, J.~Santiago and M.~Spannowsky,
  ``Constraining four-fermion operators using rare top decays,''
  JHEP {\bf 1904}, 014 (2019)
  doi:10.1007/JHEP04(2019)014
  [arXiv:1809.09624 [hep-ph]].
  %%CITATION = doi:10.1007/JHEP04(2019)014;%%
  %9 citations counted in INSPIRE as of 06 Aug 2019




\bibitem{e8}
C.~Englert, R.~Kogler, H.~Schulz and M.~Spannowsky,
  ``Higgs coupling measurements at the LHC,''
  Eur.\ Phys.\ J.\ C {\bf 76}, no. 7, 393 (2016)
  doi:10.1140/epjc/s10052-016-4227-1
  [arXiv:1511.05170 [hep-ph]].
  %%CITATION = doi:10.1140/epjc/s10052-016-4227-1;%%
  %77 citations counted in INSPIRE as of 06 Aug 2019

\bibitem{e9}
J.~A.~Aguilar-Saavedra {\it et al.},
  ``Interpreting top-quark LHC measurements in the standard-model effective field theory,''
  arXiv:1802.07237 [hep-ph].
  %%CITATION = ARXIV:1802.07237;%%
  %52 citations counted in INSPIRE as of 06 Aug 2019





\bibitem{e10}
 I.~Brivio and M.~Trott,
  ``The Standard Model as an Effective Field Theory,''
  Phys.\ Rept.\  {\bf 793}, 1 (2019)
  doi:10.1016/j.physrep.2018.11.002
  [arXiv:1706.08945 [hep-ph]].
  %%CITATION = doi:10.1016/j.physrep.2018.11.002;%%
  %82 citations counted in INSPIRE as of 06 Aug 2019

\bibitem{e11}
I.~Brivio, T.~Corbett and M.~Trott,
  ``The Higgs width in the SMEFT,''
  arXiv:1906.06949 [hep-ph].
  %%CITATION = ARXIV:1906.06949;%%
  %1 citations counted in INSPIRE as of 06 Aug 2019


%\cite{Khanpour:2017cfq}
\bibitem{Khanpour:2017cfq}
  H.~Khanpour and M.~Mohammadi Najafabadi,
  ``Constraining Higgs boson effective couplings at electron-positron colliders,''
  Phys.\ Rev.\ D {\bf 95}, no. 5, 055026 (2017)
  doi:10.1103/PhysRevD.95.055026
  [arXiv:1702.00951 [hep-ph]].
  %%CITATION = doi:10.1103/PhysRevD.95.055026;%%
  %19 citations counted in INSPIRE as of 27 Aug 2019
  
  
%\cite{Hesari:2018ssq}
\bibitem{Hesari:2018ssq}
  H.~Hesari, H.~Khanpour and M.~Mohammadi Najafabadi,
  ``Study of Higgs Effective Couplings at Electron-Proton Colliders,''
  Phys.\ Rev.\ D {\bf 97}, no. 9, 095041 (2018)
  doi:10.1103/PhysRevD.97.095041
  [arXiv:1805.04697 [hep-ph]].
  %%CITATION = doi:10.1103/PhysRevD.97.095041;%%
  %5 citations counted in INSPIRE as of 27 Aug 2019


%\cite{Khanpour:2017inb}
\bibitem{Khanpour:2017inb}
  H.~Khanpour, S.~Khatibi and M.~Mohammadi Najafabadi,
  ``Probing Higgs boson couplings in H+$\gamma$ production at the LHC,''
  Phys.\ Lett.\ B {\bf 773}, 462 (2017)
  doi:10.1016/j.physletb.2017.09.005
  [arXiv:1702.05753 [hep-ph]].
  %%CITATION = doi:10.1016/j.physletb.2017.09.005;%%
  %15 citations counted in INSPIRE as of 27 Aug 2019





%\cite{Ellis:2015sca}
\bibitem{Ellis:2015sca}
J.~Ellis and T.~You,
  ``Sensitivities of Prospective Future e+e- Colliders to Decoupled New Physics,''
  JHEP {\bf 1603}, 089 (2016)
  doi:10.1007/JHEP03(2016)089
  [arXiv:1510.04561 [hep-ph]].
  %%CITATION = doi:10.1007/JHEP03(2016)089;%%
  %54 citations counted in INSPIRE as of 27 Aug 2019


%\cite{Chiu:2017yrx}
\bibitem{Chiu:2017yrx}
  W.~H.~Chiu, S.~C.~Leung, T.~Liu, K.~F.~Lyu and L.~T.~Wang,
  ``Probing 6D operators at future e$^{?}$e$^{+}$ colliders,''
  JHEP {\bf 1805}, 081 (2018)
  doi:10.1007/JHEP05(2018)081
  [arXiv:1711.04046 [hep-ph]].
  %%CITATION = doi:10.1007/JHEP05(2018)081;%%
  %10 citations counted in INSPIRE as of 27 Aug 2019



%\cite{Ellis:2017kfi}
\bibitem{Ellis:2017kfi}
  J.~Ellis, P.~Roloff, V.~Sanz and T.~You,
  ``Dimension-6 Operator Analysis of the CLIC Sensitivity to New Physics,''
  JHEP {\bf 1705}, 096 (2017)
  doi:10.1007/JHEP05(2017)096
  [arXiv:1701.04804 [hep-ph]].
  %%CITATION = doi:10.1007/JHEP05(2017)096;%%
  %38 citations counted in INSPIRE as of 27 Aug 2019



%\cite{Brooijmans:2016vro}
\bibitem{Brooijmans:2016vro}
 G.~Brooijmans {\it et al.},
  ``Les Houches 2015: Physics at TeV colliders - new physics working group report,''
  arXiv:1605.02684 [hep-ph].
  %%CITATION = ARXIV:1605.02684;%%
  %34 citations counted in INSPIRE as of 27 Aug 2019



%\cite{Rontsch:2015una}
\bibitem{Rontsch:2015una} 
  R.~Rontsch and M.~Schulze,
  ``Probing top-Z dipole moments at the LHC and ILC,''
  JHEP {\bf 1508}, 044 (2015)
  doi:10.1007/JHEP08(2015)044
  [arXiv:1501.05939 [hep-ph]].
  %%CITATION = doi:10.1007/JHEP08(2015)044;%%
  %50 citations counted in INSPIRE as of 27 Aug 2019




%\cite{Rontsch:2014cca}
\bibitem{Rontsch:2014cca} 
  R.~Rontsch and M.~Schulze,
  ``Constraining couplings of top quarks to the Z boson in $ t\overline{t} $ + Z production at the LHC,''
  JHEP {\bf 1407}, 091 (2014)
  Erratum: [JHEP {\bf 1509}, 132 (2015)]
  doi:10.1007/JHEP09(2015)132, 10.1007/JHEP07(2014)091
  [arXiv:1404.1005 [hep-ph]].
  %%CITATION = doi:10.1007/JHEP09(2015)132, 10.1007/JHEP07(2014)091;%%
  %60 citations counted in INSPIRE as of 27 Aug 2019



\bibitem{r1}
  G.~Durieux, C.~Grojean, J.~Gu and K.~Wang,
  ``The leptonic future of the Higgs,''
  JHEP {\bf 1709}, 014 (2017)
  doi:10.1007/JHEP09(2017)014
  [arXiv:1704.02333 [hep-ph]].
  %%CITATION = doi:10.1007/JHEP09(2017)014;%%
  %51 citations counted in INSPIRE as of 27 Aug 2019

\bibitem{r2}
  A.~Buckley, C.~Englert, J.~Ferrando, D.~J.~Miller, L.~Moore, M.~Russell and C.~D.~White,
  ``Constraining top quark effective theory in the LHC Run II era,''
  JHEP {\bf 1604}, 015 (2016)
  doi:10.1007/JHEP04(2016)015
  [arXiv:1512.03360 [hep-ph]].
  %%CITATION = doi:10.1007/JHEP04(2016)015;%%
  %88 citations counted in INSPIRE as of 27 Aug 2019

\bibitem{r3}
  A.~Buckley, C.~Englert, J.~Ferrando, D.~J.~Miller, L.~Moore, M.~Russell and C.~D.~White,
  ``Global fit of top quark effective theory to data,''
  Phys.\ Rev.\ D {\bf 92}, no. 9, 091501 (2015)
  doi:10.1103/PhysRevD.92.091501
  [arXiv:1506.08845 [hep-ph]].
  %%CITATION = doi:10.1103/PhysRevD.92.091501;%%
  %72 citations counted in INSPIRE as of 27 Aug 2019

\bibitem{r4}
  M.~S.~Amjad {\it et al.},
  ``A precise determination of top quark electro-weak couplings at the ILC operating at $\sqrt{s}=500$ GeV,''
  arXiv:1307.8102 [hep-ex].
  %%CITATION = ARXIV:1307.8102;%%
  %61 citations counted in INSPIRE as of 27 Aug 2019
  
\bibitem{r5}
 M.~Baak {\it et al.} [Gfitter Group],
  ``The global electroweak fit at NNLO and prospects for the LHC and ILC,''
  Eur.\ Phys.\ J.\ C {\bf 74}, 3046 (2014)
  doi:10.1140/epjc/s10052-014-3046-5
  [arXiv:1407.3792 [hep-ph]].
  %%CITATION = doi:10.1140/epjc/s10052-014-3046-5;%%
  %518 citations counted in INSPIRE as of 27 Aug 2019


\bibitem{r6}
 I.~Brivio, Y.~Jiang and M.~Trott,
  ``The SMEFTsim package, theory and tools,''
  JHEP {\bf 1712}, 070 (2017)
  doi:10.1007/JHEP12(2017)070
  [arXiv:1709.06492 [hep-ph]].
  %%CITATION = doi:10.1007/JHEP12(2017)070;%%
  %39 citations counted in INSPIRE as of 27 Aug 2019

\bibitem{r7}
  A.~Vasquez, C.~Degrande, A.~Tonero and R.~Rosenfeld,
  ``New physics in double Higgs production at future e$^{+}$e$^{?}$ colliders,''
  JHEP {\bf 1905}, 020 (2019)
  doi:10.1007/JHEP05(2019)020
  [arXiv:1901.05979 [hep-ph]].
  %%CITATION = doi:10.1007/JHEP05(2019)020;%%



\bibitem{r8}
 D.~Atwood, S.~Bar-Shalom, G.~Eilam and A.~Soni,
  ``CP violation in top physics,''
  Phys.\ Rept.\  {\bf 347}, 1 (2001)
  doi:10.1016/S0370-1573(00)00112-5
  [hep-ph/0006032].
  %%CITATION = doi:10.1016/S0370-1573(00)00112-5;%%
  %168 citations counted in INSPIRE as of 27 Aug 2019


%\cite{Englert:2014uua}
\bibitem{r9} 
C.~Englert, A.~Freitas, M.~M.~Muhlleitner, T.~Plehn, M.~Rauch, M.~Spira and K.~Walz,
``Precision Measurements of Higgs Couplings: Implications for New Physics Scales,''
J.\ Phys.\ G {\bf 41}, 113001 (2014),
%doi:10.1088/0954-3899/41/11/113001
[arXiv:1403.7191 [hep-ph]].
%%CITATION = doi:10.1088/0954-3899/41/11/113001;%%
%132 citations counted in INSPIRE as of 14 Jun 2019


\bibitem{r10} 
  J.~A.~Aguilar-Saavedra, B.~Fuks and M.~L.~Mangano,
  ``Pinning down top dipole moments with ultra-boosted tops,''
  Phys.\ Rev.\ D {\bf 91}, 094021 (2015)
  doi:10.1103/PhysRevD.91.094021
  [arXiv:1412.6654 [hep-ph]].
  %%CITATION = doi:10.1103/PhysRevD.91.094021;%%
  %42 citations counted in INSPIRE as of 27 Aug 2019
  
  \bibitem{r11} 
  M.~Mohammadi Najafabadi,
  ``Probing of Wtb Anomalous Couplings via the tW Channel of Single Top Production,''
  JHEP {\bf 0803}, 024 (2008)
  doi:10.1088/1126-6708/2008/03/024
  [arXiv:0801.1939 [hep-ph]].
  %%CITATION = doi:10.1088/1126-6708/2008/03/024;%%
  %23 citations counted in INSPIRE as of 27 Aug 2019


%\cite{Ellis:2014jta}
\bibitem{Ellis:2014jta} 
J.~Ellis, V.~Sanz and T.~You,
  ``The Effective Standard Model after LHC Run I,''
  JHEP {\bf 1503}, 157 (2015)
  doi:10.1007/JHEP03(2015)157
  [arXiv:1410.7703 [hep-ph]].
  %%CITATION = doi:10.1007/JHEP03(2015)157;%%
  %158 citations counted in INSPIRE as of 27 Aug 2019


%\cite{Englert:2015hrx}
\bibitem{Englert:2015hrx} 
 C.~Englert, R.~Kogler, H.~Schulz and M.~Spannowsky,
  ``Higgs coupling measurements at the LHC,''
  Eur.\ Phys.\ J.\ C {\bf 76}, no. 7, 393 (2016)
  doi:10.1140/epjc/s10052-016-4227-1
  [arXiv:1511.05170 [hep-ph]].
  %%CITATION = doi:10.1140/epjc/s10052-016-4227-1;%%
  %78 citations counted in INSPIRE as of 27 Aug 2019

%\cite{Ellis:2014dva}
\bibitem{Ellis:2014dva} 
 J.~Ellis, V.~Sanz and T.~You,
  ``Complete Higgs Sector Constraints on Dimension-6 Operators,''
  JHEP {\bf 1407}, 036 (2014)
  doi:10.1007/JHEP07(2014)036
  [arXiv:1404.3667 [hep-ph]].
  %%CITATION = doi:10.1007/JHEP07(2014)036;%%
  %158 citations counted in INSPIRE as of 27 Aug 2019

%\cite{Denizli:2017pyu}
\bibitem{Denizli:2017pyu} 
  H.~Denizli and A.~Senol,
  ``Constraints on Higgs effective couplings in $H\nu \bar{\nu}$ production of CLIC at 380 GeV,''
  Adv.\ High Energy Phys.\  {\bf 2018}, 1627051 (2018)
  doi:10.1155/2018/1627051
  [arXiv:1707.03890 [hep-ph]].
  %%CITATION = doi:10.1155/2018/1627051;%%
  %5 citations counted in INSPIRE as of 27 Aug 2019


%\cite{Denizli:2019ijf}
\bibitem{Denizli:2019ijf} 
 H.~Denizli, K.~Y.~Oyulmaz and A.~Senol,
  ``Testing for observability of Higgs effective couplings in triphoton production at FCC-hh,''
  arXiv:1901.04784 [hep-ph].
  %%CITATION = ARXIV:1901.04784;%%
  
  
%\cite{Hesari:2018lzx}
\bibitem{Hesari:2018lzx} 
  H.~Hesari,
  ``Probing Higgs boson couplings in $t\bar{t}b\bar{b}$ production at the LHC,''
  arXiv:1807.04306 [hep-ph].
  %%CITATION = ARXIV:1807.04306;%%

%\cite{Dror:2015nkp}
\bibitem{ddd} 
  J.~A.~Dror, M.~Farina, E.~Salvioni and J.~Serra,
  ``Strong tW Scattering at the LHC,''
  JHEP {\bf 1601}, 071 (2016)
  doi:10.1007/JHEP01(2016)071
  [arXiv:1511.03674 [hep-ph]].
  %%CITATION = doi:10.1007/JHEP01(2016)071;%%
  %27 citations counted in INSPIRE as of 20 Oct 2019

%\cite{Hartmann:2016pil}
\bibitem{Hartmann:2016pil} 
  C.~Hartmann, W.~Shepherd and M.~Trott,
  ``The $Z$ decay width in the SMEFT: $y_t$ and $\lambda$ corrections at one loop,''
  JHEP {\bf 1703}, 060 (2017)
  doi:10.1007/JHEP03(2017)060
  [arXiv:1611.09879 [hep-ph]].
  %%CITATION = doi:10.1007/JHEP03(2017)060;%%
  %21 citations counted in INSPIRE as of 27 Aug 2019


%\cite{Fichet:2016iuo}
\bibitem{Fichet:2016iuo} 
 S.~Fichet, A.~Tonero and P.~Rebello Teles,
  ``Sharpening the shape analysis for higher-dimensional operator searches,''
  Phys.\ Rev.\ D {\bf 96}, no. 3, 036003 (2017)
  doi:10.1103/PhysRevD.96.036003
  [arXiv:1611.01165 [hep-ph]].
  %%CITATION = doi:10.1103/PhysRevD.96.036003;%%
  %5 citations counted in INSPIRE as of 27 Aug 2019

%\cite{Berthier:2015gja}
\bibitem{Berthier:2015gja} 
 L.~Berthier and M.~Trott,
  ``Consistent constraints on the Standard Model Effective Field Theory,''
  JHEP {\bf 1602}, 069 (2016)
  doi:10.1007/JHEP02(2016)069
  [arXiv:1508.05060 [hep-ph]].
  %%CITATION = doi:10.1007/JHEP02(2016)069;%%
  %78 citations counted in INSPIRE as of 27 Aug 2019

%\cite{Durieux:2018tev}
\bibitem{Durieux:2018tev} 
  G.~Durieux, M.~Perelló, M.~Vos and C.~Zhang,
  ``Global and optimal probes for the top-quark effective field theory at future lepton colliders,''
  JHEP {\bf 1810}, 168 (2018)
  doi:10.1007/JHEP10(2018)168
  [arXiv:1807.02121 [hep-ph]].
  %%CITATION = doi:10.1007/JHEP10(2018)168;%%
  %20 citations counted in INSPIRE as of 06 Jan 2020


%\cite{Artoisenet:2013puc}
\bibitem{Artoisenet:2013puc}
  P.~Artoisenet {\it et al.},
  ``A framework for Higgs characterisation,''
  JHEP {\bf 1311}, 043 (2013)
  doi:10.1007/JHEP11(2013)043
  [arXiv:1306.6464 [hep-ph]].
  %%CITATION = doi:10.1007/JHEP11(2013)043;%%
  %165 citations counted in INSPIRE as of 27 Aug 2019


%\cite{Alloul:2013naa}
\bibitem{Alloul:2013naa}
  A.~Alloul, B.~Fuks and V.~Sanz,
  ``Phenomenology of the Higgs Effective Lagrangian via FEYNRULES,''
  JHEP {\bf 1404}, 110 (2014)
  doi:10.1007/JHEP04(2014)110
  [arXiv:1310.5150 [hep-ph]].
  %%CITATION = doi:10.1007/JHEP04(2014)110;%%
  %107 citations counted in INSPIRE as of 27 Aug 2019

%\cite{Buchmuller:1985jz}
\bibitem{Buchmuller:1985jz} 
 W.~Buchmuller and D.~Wyler,
  ``Effective Lagrangian Analysis of New Interactions and Flavor Conservation,''
  Nucl.\ Phys.\ B {\bf 268}, 621 (1986).
  doi:10.1016/0550-3213(86)90262-2
  %%CITATION = doi:10.1016/0550-3213(86)90262-2;%%
  %1569 citations counted in INSPIRE as of 27 Aug 2019




%\cite{Grzadkowski:2010es}
\bibitem{Grzadkowski:2010es} 
  B.~Grzadkowski, M.~Iskrzynski, M.~Misiak and J.~Rosiek,
  ``Dimension-Six Terms in the Standard Model Lagrangian,''
  JHEP {\bf 1010}, 085 (2010)
  doi:10.1007/JHEP10(2010)085
  [arXiv:1008.4884 [hep-ph]].
  %%CITATION = doi:10.1007/JHEP10(2010)085;%%
  %934 citations counted in INSPIRE as of 27 Aug 2019





%\cite{Hagiwara:1993ck}
\bibitem{Hagiwara:1993ck} 
K.~Hagiwara, S.~Ishihara, R.~Szalapski and D.~Zeppenfeld,
  ``Low-energy effects of new interactions in the electroweak boson sector,''
  Phys.\ Rev.\ D {\bf 48}, 2182 (1993).
  doi:10.1103/PhysRevD.48.2182
  %%CITATION = doi:10.1103/PhysRevD.48.2182;%%
  %634 citations counted in INSPIRE as of 27 Aug 2019



%\cite{Buchalla:2014eca}
\bibitem{Buchalla:2014eca} 
   G.~Buchalla, O.~Cata and C.~Krause,
  ``A Systematic Approach to the SILH Lagrangian,''
  Nucl.\ Phys.\ B {\bf 894}, 602 (2015)
  doi:10.1016/j.nuclphysb.2015.03.024
  [arXiv:1412.6356 [hep-ph]].
  %%CITATION = doi:10.1016/j.nuclphysb.2015.03.024;%%
  %49 citations counted in INSPIRE as of 27 Aug 2019
    
  %\cite{Buchalla:2015wfa}
\bibitem{Buchalla:2015wfa} 
  G.~Buchalla, O.~Cata, A.~Celis and C.~Krause,
  ``Note on Anomalous Higgs-Boson Couplings in Effective Field Theory,''
  Phys.\ Lett.\ B {\bf 750}, 298 (2015)
  doi:10.1016/j.physletb.2015.09.027
  [arXiv:1504.01707 [hep-ph]].
  %%CITATION = doi:10.1016/j.physletb.2015.09.027;%%
  %37 citations counted in INSPIRE as of 09 Aug 2019
  
  %\cite{Contino:2013kra}
\bibitem{Contino:2013kra} 
  R.~Contino, M.~Ghezzi, C.~Grojean, M.~Muhlleitner and M.~Spira,
  ``Effective Lagrangian for a light Higgs-like scalar,''
  JHEP {\bf 1307}, 035 (2013)
  doi:10.1007/JHEP07(2013)035
  [arXiv:1303.3876 [hep-ph]].
  %%CITATION = doi:10.1007/JHEP07(2013)035;%%
  %326 citations counted in INSPIRE as of 09 Aug 2019
  



%\cite{Englert:2017dev}
\bibitem{Englert:2017dev} 
  C.~Englert and M.~Russell,
  ``Top quark electroweak couplings at future lepton colliders,''
  Eur.\ Phys.\ J.\ C {\bf 77}, no. 8, 535 (2017)
  doi:10.1140/epjc/s10052-017-5095-z
  [arXiv:1704.01782 [hep-ph]].
  %%CITATION = doi:10.1140/epjc/s10052-017-5095-z;%%
  %18 citations counted in INSPIRE as of 21 Aug 2019
 

\bibitem{cms1} 
  A.~M.~Sirunyan {\it et al.} [CMS Collaboration],
  ``Measurement of the cross section for top quark pair production in association with a W or Z boson in proton-proton collisions at $\sqrt{s} =$ 13 TeV,''
  JHEP {\bf 1808}, 011 (2018)
  doi:10.1007/JHEP08(2018)011
  [arXiv:1711.02547 [hep-ex]].
  %%CITATION = doi:10.1007/JHEP08(2018)011;%%
  %65 citations counted in INSPIRE as of 02 Jan 2020


%\cite{CMS:2019too}
\bibitem{cms2} 
  [CMS Collaboration],
  ``Measurement of top quark pair production in association with a Z boson in proton-proton collisions at $\sqrt{s}=$ 13 TeV,''
  arXiv:1907.11270 [hep-ex].
  %%CITATION = ARXIV:1907.11270;%%
  %10 citations counted in INSPIRE as of 02 Jan 2020


%\cite{Hartland:2019bjb}
\bibitem{gf1} 
  N.~P.~Hartland, F.~Maltoni, E.~R.~Nocera, J.~Rojo, E.~Slade, E.~Vryonidou and C.~Zhang,
  ``A Monte Carlo global analysis of the Standard Model Effective Field Theory: the top quark sector,''
  JHEP {\bf 1904}, 100 (2019)
  doi:10.1007/JHEP04(2019)100
  [arXiv:1901.05965 [hep-ph]].
  %%CITATION = doi:10.1007/JHEP04(2019)100;%%
  %40 citations counted in INSPIRE as of 02 Jan 2020

%\cite{Brivio:2019ius}
\bibitem{gf2} 
  I.~Brivio, S.~Bruggisser, F.~Maltoni, R.~Moutafis, T.~Plehn, E.~Vryonidou, S.~Westhoff and C.~Zhang,
  ``O new physics, where art thou? A global search in the top sector,''
  arXiv:1910.03606 [hep-ph].
  %%CITATION = ARXIV:1910.03606;%%
  %4 citations counted in INSPIRE as of 02 Jan 2020

\bibitem{gf3} 
  J.~Ellis, C.~W.~Murphy, V.~Sanz and T.~You,
  %``Updated Global SMEFT Fit to Higgs, Diboson and Electroweak Data,''
  JHEP {\bf 1806}, 146 (2018)
  doi:10.1007/JHEP06(2018)146
  [arXiv:1803.03252 [hep-ph]].
  %%CITATION = doi:10.1007/JHEP06(2018)146;%%
  %80 citations counted in INSPIRE as of 08 Jan 2020


%\cite{Alwall:2011uj}
\bibitem{Alwall:2011uj}
 J.~Alwall, M.~Herquet, F.~Maltoni, O.~Mattelaer and T.~Stelzer,
  ``MadGraph 5 : Going Beyond,''
  JHEP {\bf 1106}, 128 (2011)
  doi:10.1007/JHEP06(2011)128
  [arXiv:1106.0522 [hep-ph]].
  %%CITATION = doi:10.1007/JHEP06(2011)128;%%
  %2890 citations counted in INSPIRE as of 27 Aug 2019


%\cite{Alwall:2014bza}
\bibitem{Alwall:2014bza}
 J.~Alwall, C.~Duhr, B.~Fuks, O.~Mattelaer, D.~G.~Ozturk and C.~H.~Shen,
  ``Computing decay rates for new physics theories with FeynRules  and MadGraph 5\_aMC@NLO,''
  Comput.\ Phys.\ Commun.\  {\bf 197}, 312 (2015)
  doi:10.1016/j.cpc.2015.08.031
  [arXiv:1402.1178 [hep-ph]].
  %%CITATION = doi:10.1016/j.cpc.2015.08.031;%%
  %61 citations counted in INSPIRE as of 27 Aug 2019






%\cite{Alwall:2014hca}
\bibitem{Alwall:2014hca}
J.~Alwall {\it et al.},
  ``The automated computation of tree-level and next-to-leading order differential cross sections, and their matching to parton shower simulations,''
  JHEP {\bf 1407}, 079 (2014)
  doi:10.1007/JHEP07(2014)079
  [arXiv:1405.0301 [hep-ph]].
  %%CITATION = doi:10.1007/JHEP07(2014)079;%%
  %3853 citations counted in INSPIRE as of 27 Aug 2019





%\cite{Alloul:2013bka}
\bibitem{Alloul:2013bka} 
  A.~Alloul, N.~D.~Christensen, C.~Degrande, C.~Duhr and B.~Fuks,
  ``FeynRules  2.0 - A complete toolbox for tree-level phenomenology,''
  Comput.\ Phys.\ Commun.\  {\bf 185}, 2250 (2014)
  doi:10.1016/j.cpc.2014.04.012
  [arXiv:1310.1921 [hep-ph]].
  %%CITATION = doi:10.1016/j.cpc.2014.04.012;%%
  %1178 citations counted in INSPIRE as of 27 Aug 2019

%\cite{Degrande:2011ua}
\bibitem{Degrande:2011ua} 
 C.~Degrande, C.~Duhr, B.~Fuks, D.~Grellscheid, O.~Mattelaer and T.~Reiter,
  ``UFO - The Universal FeynRules Output,''
  Comput.\ Phys.\ Commun.\  {\bf 183}, 1201 (2012)
  doi:10.1016/j.cpc.2012.01.022
  [arXiv:1108.2040 [hep-ph]].
  %%CITATION = doi:10.1016/j.cpc.2012.01.022;%%
  %653 citations counted in INSPIRE as of 27 Aug 2019

\bibitem{mlm}
 M.~L.~Mangano, M.~Moretti, F.~Piccinini and M.~Treccani,
  ``Matching matrix elements and shower evolution for top-quark production in hadronic collisions,''
  JHEP {\bf 0701}, 013 (2007)
  doi:10.1088/1126-6708/2007/01/013
  [hep-ph/0611129].
  %%CITATION = doi:10.1088/1126-6708/2007/01/013;%%
  %673 citations counted in INSPIRE as of 02 Jan 2020



%\cite{Tanabashi:2018oca}
\bibitem{Tanabashi:2018oca}
  M.~Tanabashi {\it et al.} [Particle Data Group],
  ``Review of Particle Physics,''
  Phys.\ Rev.\ D {\bf 98}, no. 3, 030001 (2018).
  doi:10.1103/PhysRevD.98.030001
  %%CITATION = doi:10.1103/PhysRevD.98.030001;%%
  %2315 citations counted in INSPIRE as of 27 Aug 2019



%\cite{Sjostrand:2014zea}
\bibitem{Sjostrand:2014zea}
T.~Sjostrand {\it et al.},
  ``An Introduction to PYTHIA 8.2,''
  Comput.\ Phys.\ Commun.\  {\bf 191}, 159 (2015)
  doi:10.1016/j.cpc.2015.01.024
  [arXiv:1410.3012 [hep-ph]].
  %%CITATION = doi:10.1016/j.cpc.2015.01.024;%%
  %1784 citations counted in INSPIRE as of 27 Aug 2019






%\cite{Sjostrand:2007gs}
\bibitem{Sjostrand:2007gs}
 T.~Sjostrand, S.~Mrenna and P.~Z.~Skands,
  ``A Brief Introduction to PYTHIA 8.1,''
  Comput.\ Phys.\ Commun.\  {\bf 178}, 852 (2008)
  doi:10.1016/j.cpc.2008.01.036
  [arXiv:0710.3820 [hep-ph]].
  %%CITATION = doi:10.1016/j.cpc.2008.01.036;%%
  %4513 citations counted in INSPIRE as of 27 Aug 2019



%\cite{deFavereau:2013fsa}
\bibitem{deFavereau:2013fsa}
  J.~de Favereau {\it et al.} [DELPHES 3 Collaboration],
  ``DELPHES 3, A modular framework for fast simulation of a generic collider experiment,''
  JHEP {\bf 1402}, 057 (2014)
  doi:10.1007/JHEP02(2014)057
  [arXiv:1307.6346 [hep-ex]].
  %%CITATION = doi:10.1007/JHEP02(2014)057;%%
  %1304 citations counted in INSPIRE as of 27 Aug 2019



%\cite{Behnke:2013lya}
\bibitem{Behnke:2013lya}
  T.~Behnke {\it et al.},
  ``The International Linear Collider Technical Design Report - Volume 4: Detectors,''
  arXiv:1306.6329 [physics.ins-det].
  %%CITATION = ARXIV:1306.6329;%%
  %449 citations counted in INSPIRE as of 27 Aug 2019




%\cite{Cacciari:2008gp}
\bibitem{Cacciari:2008gp}
 M.~Cacciari, G.~P.~Salam and G.~Soyez,
  ``The anti-$k_t$ jet clustering algorithm,''
  JHEP {\bf 0804}, 063 (2008)
  doi:10.1088/1126-6708/2008/04/063
  [arXiv:0802.1189 [hep-ph]].
  %%CITATION = doi:10.1088/1126-6708/2008/04/063;%%
  %6111 citations counted in INSPIRE as of 27 Aug 2019



%\cite{Cacciari:2011ma}
\bibitem{Cacciari:2011ma}
 M.~Cacciari, G.~P.~Salam and G.~Soyez,
  ``FastJet User Manual,''
  Eur.\ Phys.\ J.\ C {\bf 72}, 1896 (2012)
  doi:10.1140/epjc/s10052-012-1896-2
  [arXiv:1111.6097 [hep-ph]].
  %%CITATION = doi:10.1140/epjc/s10052-012-1896-2;%%
  %2989 citations counted in INSPIRE as of 27 Aug 2019
  


%\cite{Hocker:2007ht}
\bibitem{Hocker:2007ht} 
A.~Hocker {\it et al.},
``TMVA - Toolkit for Multivariate Data Analysis,''
PoS ACAT {\bf }, 040 (2007)
[physics/0703039 [PHYSICS]].
%%CITATION = PHYSICS/0703039;%%
%731 citations counted in INSPIRE as of 29 Sep 2016

%\cite{Stelzer:2008zz}
\bibitem{Stelzer:2008zz} 
J.~Stelzer, A.~Hocker, P.~Speckmayer and H.~Voss,
``Current developments in TMVA: An outlook to TMVA4,''
PoS ACAT {\bf 08}, 063 (2008).
%%CITATION = POSCI,ACAT08,063;%%
%2 citations counted in INSPIRE as of 29 Sep 2016

%\cite{Therhaag:2009dp}
\bibitem{Therhaag:2009dp} 
J.~Therhaag [TMVA Core Developer Team Collaboration],
``TMVA: Toolkit for multivariate data analysis,''
AIP Conf.\ Proc.\  {\bf 1504}, 1013 (2009).
%%doi:10.1063/1.4771869
%%CITATION = doi:10.1063/1.4771869;%%
%16 citations counted in INSPIRE as of 29 Sep 2016

%\cite{Speckmayer:2010zz}
\bibitem{Speckmayer:2010zz} 
P.~Speckmayer, A.~Hocker, J.~Stelzer and H.~Voss,
``The toolkit for multivariate data analysis, TMVA 4,''
J.\ Phys.\ Conf.\ Ser.\  {\bf 219}, 032057 (2010).
%%doi:10.1088/1742-6596/219/3/032057
%%CITATION = doi:10.1088/1742-6596/219/3/032057;%%
%50 citations counted in INSPIRE as of 29 Sep 2016

%\cite{Therhaag:2010zz}
\bibitem{Therhaag:2010zz} 
J.~Therhaag,
``TMVA Toolkit for multivariate data analysis in ROOT,''
PoS ICHEP {\bf 2010}, 510 (2010).
%%CITATION = POSCI,ICHEP2010,510;%%
%4 citations counted in INSPIRE as of 29 Sep 2016


 
\bibitem{Giudice:2007fh} 
  G.~F.~Giudice, C.~Grojean, A.~Pomarol and R.~Rattazzi,
  ``The Strongly-Interacting Light Higgs,''
 JHEP {\bf 0706}, 045 (2007)
  doi:10.1088/1126-6708/2007/06/045
  [hep-ph/0703164].
  %%CITATION = doi:10.1088/1126-6708/2007/06/045;%%
  %828 citations counted in INSPIRE as of 27 Aug 2019


\end{thebibliography}
\end{document}